\documentclass[11pt,a4paper]{article}
\usepackage{amsmath,amssymb,hyperref,bm,bbm}
\usepackage[utf8]{inputenc}
\numberwithin{equation}{section}
\allowdisplaybreaks[3]
\usepackage{braket}
\usepackage{subcaption}

\begin{document}
\begin{titlepage}

\renewcommand{\thefootnote}{\fnsymbol{footnote}}
\begin{flushright}
\begin{tabular}{l}
YITP-21-144\\
\end{tabular}
\end{flushright}

\vfill
\begin{center}


\noindent{\large \textbf{FZZ-triality and large $\mathcal{N}=4$ super Liouville theory} }

 \medskip

\noindent{\large \textbf{}}

\vspace{1.5cm}

\noindent{Thomas Creutzig$^{a}$\footnote{E-mail: creutzig@ualberta.ca} and Yasuaki Hikida$^b$\footnote{E-mail: yhikida@yukawa.kyoto-u.ac.jp}}
\bigskip

\vskip .6 truecm
\centerline{\it $^a$Department of Mathematical and Statistical Sciences, University of Alberta,} \centerline{\it Edmonton, Alberta T6G 2G1, Canada}
\medskip
\centerline{\it $^b$Center for Gravitational Physics, Yukawa Institute for Theoretical Physics,}
\centerline{\it  Kyoto University, Kyoto 606-8502, Japan}

\end{center}

\vfill
\vskip 0.5 truecm
\begin{abstract}

We examine dualities of two dimensional conformal field theories by applying the methods developed in previous works. We first derive the duality between $SL(2|1)_k/(SL(2)_k \otimes U(1))$ coset and Witten's cigar model or sine-Liouville theory. The latter two models are Fateev-Zamolodchikov-Zamolodchikov (FZZ-)dual to each other, hence the relation of the three models is named  FZZ-triality. These results are used to study correlator correspondences between large $\mathcal{N}=4$ super Liouville theory and a coset of the form $Y(k_1,k_2)/SL(2)_{k_1 +k_2}$, where $Y(k_1 , k_2)$ consists of two  $SL(2|1)_{k_i}$ and free bosons or equivalently two $U(1)$ cosets of $D(2,1;k_i -1)$ at level one. These correspondences are a main result of this paper.  The FZZ-triality acts as a seed of the correspondence, which in particular implies a hidden $SL(2)_{k'}$ in $SL(2|1)_k$ or $D(2,1 ; k-1)_1$.  The relation of levels is $k' -1 = 1/(k-1)$. We also construct boundary actions in sine-Liouville theory as another use of the FZZ-triality. Furthermore, we generalize the FZZ-triality to the case with $SL(n|1)_k/(SL(n)_k \otimes U(1))$ for arbitrary $n>2$.

\end{abstract}
\vfill
\vskip 0.5 truecm

\setcounter{footnote}{0}
\renewcommand{\thefootnote}{\arabic{footnote}}
\end{titlepage}

\newpage

\tableofcontents

\section{Introduction and summary}

We examine dualities in two dimensional conformal field theories by deriving correlator correspondences of primary operators. Combined with the match of symmetry algebras, we can deduce the equivalence of dual theories.  
In general, dualities are quite useful since it often happens that some features of a theory are easily captured by dual one. In particular, a strong/weak duality helps us to examine strong coupling phenomena from a dual theory in a tractable regime.
In previous works \cite{Hikida:2008pe,Creutzig:2010bt,Creutzig:2020cmn,Creutzig:2021cyl,Creutzig:2021ykz}, correlator correspondences for several important dualities have been derived by applying the reduction method from $SL(2)$ Wess-Zumino-Novikov-Witten (WZNW) model to Liouville field theory \cite{Ribault:2005wp,Ribault:2005ms,Hikida:2007tq} and its generalizations \cite{Hikida:2007sz,Creutzig:2011qm,Creutzig:2015hla,Creutzig:2020ffn}. 
The results in the previous works may be regarded as conformal field theory realizations of dualities of corner vertex operator algebras (VOAs) conjectured by  Gaiotto and Rap\v{c}\'ak via brane junctions in superstring theory \cite{Gaiotto:2017euk} and proven in \cite{Creutzig:2020zaj,Creutzig:2021dda}. In this paper, we examine more dualities, which are not be directly related to the Gaiotto-Rap\v{c}\'ak dualities. In particular, we derive correlator correspondences between the large $\mathcal{N}=4$ super Liouville theory and a coset model, where the agreement of symmetry algebra was proven in \cite{Creutzig:2019kro}.

Let us put this new duality into context:
An important (strong/weak) duality is the one conjectured by Fateev-Zamolodchikov-Zamolodchikov (FZZ) \cite{FZZ} and proven in \cite{Hikida:2008pe}. The  FZZ-duality is an equivalence between two dimensional cigar model described by the coset  \cite{Witten:1991yr} 
\begin{align}
\frac{SL(2)_k}{U(1)} \label{cigar}
\end{align}
and sine-Liouville theory, see, e.g.,  \cite{Kazakov:2000pm} for details. Its supersymmetric version was proven  in \cite{Hori:2001ax}  as a mirror symmetry and in \cite{Creutzig:2010bt} by the method similar to \cite{Hikida:2008pe}. The supersymmetric version is useful to examine, e.g., singular Calabi-Yau geometry \cite{Ooguri:1995wj}. Two series of generalization of the FZZ-duality have been analyzed in \cite{Creutzig:2020cmn,Creutzig:2021cyl,Creutzig:2021ykz}. A generalization is given by replacing the coset \eqref{cigar} by a higher rank one \cite{Creutzig:2020cmn,Creutzig:2021ykz}
\begin{align}
\frac{SL(n+1)_k}{SL(n)_k \otimes U(1)} \, .
\end{align}
The coset appears as conformal field theory dual of higher spin (super-)gravity \cite{Gaberdiel:2010pz,Creutzig:2011fe}, thus the duality should be useful in that context as well. Another generalization is the free boson, i.e. $U(1)$,  coset of subregular W-algebras of $\mathfrak{sl}(n)$ \cite{Creutzig:2021cyl}. Arguably the most important example is the duality between $\mathfrak{sl}(n)$ Toda field theory and the diagonal coset
\begin{align}
 \frac{SL(n)_k \otimes SL(n)_{-1}}{SL(n)_{k-1}} \label{Lcoset}
\end{align}
 analyzed in \cite{Creutzig:2021ykz}. The duality may be regarded as an analytic continuation of coset realization of W$_n$-minimal model proven only rather recently in \cite{Arakawa:2018iyk}.

We have derived correlator correspondences by applying the reduction methods from $SL(n)$ WZNW model to $\mathfrak{sl}(n)$ Toda field theory developed particularly in \cite{Hikida:2007tq,Creutzig:2020ffn}.  An important process  is to adopt a simple first order formulation of coset models. Such a formulation was proposed in \cite{Gerasimov:1989mz,Kuwahara:1989xy} and established in our recent paper \cite{Creutzig:2021ykz}.  A main idea is to express the denominator and numerator algebras of coset models by Wakimoto free field realizations with free bosons and $(\beta, \gamma)$-systems \cite{Wakimoto:1986gf}. The proposal of \cite{Gerasimov:1989mz,Kuwahara:1989xy} is to construct field space by using orthogonal free bosons and removing some sets of $(\beta, \gamma)$-systems. We can indeed show that the central charge of the reduced theory matches with that of the original coset model. In  \cite{Creutzig:2021ykz}, we have not only derived the procedure but also established a way to obtain proper interaction terms by applying the BRST formulation of coset models \cite{Gawedzki:1988nj,Karabali:1988au,Karabali:1989dk,Hwang:1993nc} and Kugo-Ogima method \cite{Kugo:1979gm}.  In particular, it can be shown that the coset algebra obtained in this way is isomorphic to the one by Goddard-Kent-Olive (GKO) construction \cite{Goddard:1986ee}. Fortunately, the first order formulation of coset models is general enough to apply to more theories, in particular the problem of this work.

In this paper, we first examine correlation functions of the coset
\begin{align}
	\frac{SL(2|1)_k}{SL(2)_k \otimes U(1)} \, .\label{trialitycoset}
\end{align}
As suggested by Gaiotto-Rap\v{c}\'ak dualities \cite{Gaiotto:2017euk} (see also \cite{Bowcock:1998kw}), the coset 
is conjecturally dual to the cigar model described by \eqref{cigar} with level $k'$
\begin{align}
	k ' - 1 = \frac{1}{k - 1} \label{levelrel}
\end{align}
and sine-Liouville theory as well. 
We derive correlator correspondences of the ``FZZ-triality.'' 
We next derive correlator correspondences between a diagonal coset
\begin{align}
	\frac{SL(2)_{k_1} \otimes SL(2)_{k_2}}{SL(2)_{k_1 + k_2}}
	\label{FScoset}
\end{align}
and a theory with a $\mathfrak{d}(2,1;-\psi)$-structure \cite{Feigin:2001yq}. 
This can be regarded as a simplified version of our main problem with large $\mathcal{N}=4$ super Liouville theory.
In fact this  theory also appears as a coset of large $\mathcal{N}=4$ super Liouville theory \cite{Creutzig:2019kro}.

We then move to the main problem of this paper.
The coset realization of large $\mathcal{N}=4$ superconformal algebra \cite{Sevrin:1988ew,Schoutens:1988ig}%
\footnote{There are two types of large $\mathcal{N}=4$ superconformal algebra, which are often called as linear and non-linear ones. In this paper, we only deal with non-linear one.}
was provided in \cite{Creutzig:2019kro}. In terms of conformal field theory, it corresponds to a duality between a coset model and large $\mathcal{N}=4$ super Liouville theory.
The coset model takes the form \cite{Creutzig:2019kro}
\begin{align}
\frac{Y(k_1 , k_2)}{SL(2)_{k_1 + k_2}} \, , \label{cosetY}
\end{align}
where the numerator $Y(k_1 ,k_2)$ consists of $SL(2|1)_{k_1},SL(2|1)_{k_2}$ and free bosons or two $U(1)$-cosets of $D(2,1; - \psi)$ with $\psi = 1 - k_1 , 1-k_2$ at level one.
The two affine superalgebras have subalgebras $\mathfrak{sl}(2)_{k_1},\mathfrak{sl}(2)_{k_2}$ and the coset is constructed by gauging the diagonal $\mathfrak{sl}(2)_{k_1 + k_2}$. The large $\mathcal{N}=4$ superconformal algebra includes two affine $\mathfrak{sl}(2)$ subalgebras with levels $k_1 ' , k_2 '$, where the relation of levels is
$
k_1 ' - 1 = 1/( k_1 - 1 ) , k_2 ' - 1 = 1 / ( k_2  - 1 )
$
as in \eqref{levelrel}. These two affine $\mathfrak{sl}(2)_{k'}$ are hidden in $\mathfrak{sl}(2|1)_k$ or $\mathfrak{d}(2,1;k-1)$ \cite{Bowcock:1999uy,Creutzig:2017uxh,Creutzig:2018ltv}.
This implies that the FZZ-triality is a key relation for the duality between the coset \eqref{cosetY} and the large $\mathcal{N}=4$ super Liouville theory.

The large $\mathcal{N}=4$ superconformal algebra is important particularly in the context of holography. For instance, conformal field theory dual of superstrings on AdS$_3 \times $S$^3 \times $S$^3 \times$S$^1$ has the large $\mathcal{N}=4$ superconformal algebra as its symmetry \cite{Elitzur:1998mm,deBoer:1999gea,Gukov:2004ym,Eberhardt:2017fsi,Eberhardt:2017pty}, see also \cite{Tong:2014yna}.
Moreover, the Wolf space model possesses the large $\mathcal{N}=4$ superconformal symmetry  \cite{Sevrin:1988ps,VanProeyen:1989me,Sevrin:1989ce} and it was proposed to be dual to a higher spin gravity in  \cite{Gaberdiel:2013vva}. We expect that our analysis helps us to reveal some important aspects of holography. The large $\mathcal{N}=4$ superconformal algebra itself deserves further study as well.
Unitary representations were examined in \cite{Gunaydin:1988re,Petersen:1989zz,Petersen:1989pp}, and it would be interesting to investigate the spectrum from the coset viewpoints as well.

In this paper, we mainly consider correlation functions on worldsheet of sphere topology. 
It is an important problem to extend the analysis to Riemann surfaces of higher genus $g >0$.
The original FZZ-duality was extended to higher genus worldsheet in \cite{Hikida:2008pe}, and it might not be so difficult to do so more generally. However, it would be more involved to examine Riemann surfaces with boundaries. In \cite{Creutzig:2010bt}, we have shown the equivalence of boundary correlation functions for D1-branes in the cigar model \eqref{cigar} and those for D2-branes in sine-Liouville theory. We have also analyzed fermionic FZZ-duality.
In this paper, we extend the FZZ-triality to the case with boundary.
The point is that there is a somehow understood procedure to obtain actions of boundary WZNW theories on supergroups \cite{Creutzig:2008ag, Creutzig:2010zp, Creutzig:2008ek, Creutzig:2010ne}. Using in particular \cite{Creutzig:2010zp} 
we show that the FZZ-triality is useful to obtain boundary actions of sine-Liouville theory even for D1-branes and its supersymmetric counterparts form the coset \eqref{trialitycoset}.

\subsection{Organization of the paper}

The paper is organized as follows.
In the next section, we collect mathematical facts underlying the current works on dualities of two dimensional conformal field theories.
In section \ref{sec:1storder}, we review the first order formulation of coset models developed in our previous paper \cite{Creutzig:2021ykz}. In particular, we closely examine the simplest but non-trivial example of the diagonal coset \eqref{Lcoset} with $n=2$ and relate it to Liouville field theory.
In section \ref{sec:coset1}, we derive correlator correspondences between the coset \eqref{trialitycoset} and sine-Liouville theory or the cigar model \eqref{cigar} by applying the first order formulation explained in section \ref{sec:1storder}. We further consider its supersymmetric versions by adding a complex fermion.
In section \ref{sec:HD}, we derive correlator correspondences between a diagonal coset \eqref{FScoset}
 and a theory with a $\mathfrak{d}(2,1;-\psi)$-structure \cite{Feigin:2001yq}. 
 In section \ref{sec:largeN4}, we investigate the coset \eqref{cosetY}. We first realize $D(2,1;-\psi)$ at level one as a coset of $SL(2|1)$ and free bosons. We then construct $Y(k_1,k_2)$ in the numerator of the coset \eqref{cosetY} by making use of the expressions. Applying the first order formulation of \cite{Creutzig:2021ykz} and the reduction methods of \cite{Hikida:2008pe}, we rewrite the $N$-point functions of the coset  \eqref{cosetY}  by those of a different theory. Applying the fermionic FZZ-duality, we show that the theory obtained in this way is indeed the large $\mathcal{N}=4$ super Liouville theory, which corresponds to a free field realization of  large $\mathcal{N}=4$  superconformal algebra in \cite{Ito:1992nq}. We also check that a coset of the theory reduces to the model analyzed in section \ref{sec:HD}.
We consider the case with additional fermions as well. In section \ref{sec:boundary}, we construct boundary actions for branes in sine-Liouville theory and $\mathcal{N}=2$ super Liouville theory by making use of boundary FZZ-triality.
In appendix \ref{sec:conv}, we summarize our conventions for generators of Lie superalgebras $\mathfrak{sl}(2|1)$ and $\mathfrak{sl}(n|1)$ with $n > 2$.
In appendix \ref{sec:OPEs}, we explicitly write down the operator product expansions (OPEs) for generators of large $\mathcal{N}=4$ superconformal algebra.
In appendix \ref{sec:sl2bdry}, we obtain classical boundary actions for B-branes and A-branes in $SL(2|1)$ WZNW model by following the analysis for $OSP(1|2)$ WZNW model in \cite{Creutzig:2010zp}.
In appendix \ref{sec:gtriality}, we generalize the FZZ-triality analyzed in section \ref{sec:coset1} and reduce the coset 
\begin{align}
\frac{SL(n|1)_k}{SL(n)_k \otimes U(1)} \label{hrcoset}
\end{align}
with $n > 2$ to a theory with an $\mathfrak{sl}(n|1)$-structure by applying the first order formulation reviewed in section \ref{sec:1storder}.

\section{Dualities}
\label{sec:Dualities}

$S$-duality of four dimensional $\mathcal N = 4$ supersymmetric GL-twisted theories is closely related to the quantum geometric Langlands correspondence as well as dualities of vertex algebras. Both are mathematical cousins of dualities in two dimensional conformal field theory. The quantum geometric Langlands correspondence is concerned with twisted $D$-modules, which in turn arise as spaces of conformal blocks. Vertex algebras are the symmetry algebras of conformal field theory. 

We recall some results of \cite{Creutzig:2017uxh}.
Fix a compact Lie group $G$, the gauge group. Let $\psi$ be the coupling of the gauge theory, a complex number. One then considers three dimensional boundary conditions. If two such boundary conditions, say $\mathcal B_1$ and $\mathcal B_2$ intersect in a two dimensional corner, then this corner usually supports a vertex operator algebra. This VOA depends on the choice of boundary condition and on $G$. Line defects can end on vertex operators associated to modules of the VOA at the corner. The type of modules is determined by the boundary conditions $\mathcal B_1$ and $\mathcal B_2$.
The duality group is the modular group $PSL(2, \mathbb Z)$. Let $g$ be an element of $PSL(2, \mathbb Z)$.  It acts on $\psi$ via M\"obius transformation and acts on boundary conditions in a certain way, see section 2.3 of \cite{Creutzig:2017uxh}. It acts on VOAs via a certain convolution, that is it maps the VOA $V$ to a certain BRST-cohomology of $V \otimes K_g$ with $K_g$ a very special kernel VOA. In particular duality is not an isomorphism of VOAs but only a correspondence. The idea however is that this correspondence respects line defects. This should be visible in the VOA setting that VOAs that are related via duality have equivalent categories of modules. Geometrically this means that one expects to have equivalent spaces of conformal blocks, exactly as the quantum geometric Langlands correspondence predicts. From the conformal field theory point of view this means that one would like to have dualities between correlation functions of two theories whose underlying VOAs are dual. The idea is that both dualities of conformal field theories and equivalences of vertex algebra tensor categories can be explained using the kernel VOAs. The  FZZ-trialities exactly correspond to such duality operations, while  our coset realization of $\mathcal N = 4$ super Liouville theory is a low rank example of a somehow more complicated story that is under current investigation.

The kernel VOAs are of the form
\begin{equation}
A^n[\mathfrak g, \mathfrak g', \psi] = \bigoplus_{\lambda \in P^+ \cap L} V^\psi(\lambda) \otimes V^{\phi}(\tau(\lambda))
\end{equation}
with 
\begin{equation}
\frac{1}{\ell_{\mathfrak g} \psi} + \frac{1}{\ell_{\mathfrak g'} \phi} = n 
\end{equation}
and $\mathfrak g'$ is a Lie (super)algebra related to $\mathfrak g$. In most cases, the two coincide, but for example if $\mathfrak g$ is of type $\mathfrak{so}_{2m+1}$ and $n$ is odd, then $\mathfrak g'=\mathfrak{osp}_{1|2m}$. $L$ is a suitable sublattice of the weight lattice $P$ of $\mathfrak g$. $P^+$ is the set of dominant weights, $n$ is an integer (usually positive), and $\psi, \phi$ are the levels shifted by the respective dual Coxeter numbers $h^\vee_{\mathfrak g}, h^\vee_{\mathfrak g'}$.  $\tau$ is a suitable map from weights of $\mathfrak g$ to $\mathfrak g'$. Finally $\ell_{\mathfrak g}$ is the lacety of $\mathfrak g$ (and for $\mathfrak{osp}_{1|2n}$ it is two).
The existence of these VOAs has been conjectured in \cite{Creutzig:2017uxh} and proven in \cite{Creutzig:2019psu, Moriwaki:2021epl}.

The simplest example is 
\begin{equation}
A^n[\mathfrak{gl}_1, \mathfrak{gl}_1, \psi] = \bigoplus_{m \in \mathbb Z} \pi^\psi_m \otimes \pi^\phi_m \cong V_{\sqrt{n}\mathbb Z} \otimes \pi
\end{equation}
which is just the lattice VOA $V_{\sqrt{n}\mathbb Z}$ times a free boson $\pi$. Here $\pi^\psi_n$ denotes the Fock module of highest-weight $n$ of a free boson of level $\psi$.  

The general idea is that one has two VOAs, $V^{-\psi}$ and $W^\phi$, such that $V^{-\psi}$ has an affine VOA $\mathfrak{g}_{-\psi - h^\vee_{\mathfrak g}}$ of $\mathfrak g$ at level $-\psi - h^\vee_{\mathfrak g}$ as subalgebra and $W^\phi$ has an affine VOA $\mathfrak{g'}_{\phi - h^\vee_{\mathfrak g'}}$ of $\mathfrak g'$ at level $-\phi - h^\vee_{\mathfrak g'}$ as subalgebra. Assume that these two VOAs are dual in the sense that their cosets coincide 
\begin{equation}
\frac{V^{-\psi}}{\mathfrak{g}_{-\psi - h^\vee_{\mathfrak g}}} =  \frac{W^\phi}{\mathfrak{g'}_{\phi - h^\vee_{\mathfrak g'}}} \, ,
\end{equation}
the main examples of isomorphisms of such cosets are provided by Gaiotto-Rap\v{c}\'ak triality \cite{Gaiotto:2017euk} proven in \cite{Creutzig:2020zaj, Creutzig:2021dda}.
Let $H^{\text{BRST}}_{\mathfrak g}( \ \bullet \ )$ denote the semi-infinite Lie algebra cohomology of $\widehat{\mathfrak{g}}$, relative to $\mathfrak g$ \cite{Frenkel:1986dg}. 
It satisfies 
\begin{equation}
H^{\text{BRST}}_{\mathfrak g}( V^{-\psi}(\lambda) \otimes V^{\psi}(\mu)) = \begin{cases} \mathbb C & \quad \text{if} \ \mu = \lambda^*   \, ,  \\ 0 & \quad \text{else}  \, .\end{cases}
\end{equation}
The proposal  of \cite{Creutzig:2020zaj, Creutzig:2021dda} is that the Gaiotto-Rap\v{c}\'ak dualities extend to the relation
\begin{equation}\label{BRST-VOA}
H^{\text{BRST}}_{\mathfrak g}(V^{-\psi} \otimes A^1[\mathfrak g, \mathfrak g', \psi]) =  W^\phi. 
\end{equation}
See \cite{Creutzig:2021bmz} for the mathematics of this statement for $\mathfrak g = \mathfrak{gl}_1$ and \cite{Creutzig:2022yvr} for the general case. 
The simplest example is $\mathfrak g = \mathfrak{g'} = \mathfrak{gl}_1$ and $V^{-\psi}$ the WZNW theory of $SL(2)$ at level $-\psi - 2$ and $W^\phi$ the $\mathcal N=2$ superconformal algebra at same central charge as the WZNW theory. In this case the kernel VOA is just a pair of free fermions times a free boson and our procedure is exactly the Kazama-Suzuki coset realization of the super conformal algebra \cite{Kazama:1988uz,Kazama:1988qp}. The corresponding duality of conformal field theories is the well established $H_3^+$ model to $\mathcal N=2$ super Liouville duality \cite{Hori:2001ax,Creutzig:2010bt}.
The natural generalization is Feigin-Semikhatov's duality between subregular $W$-algebras and principal $W$-superalgebras \cite{Feigin:2004wb}, proven in \cite{Creutzig:2020vbt}.  

This means if we take a module $M$ of $V^{-\psi}$, then 
$H^{\text{BRST}}_{\mathfrak g}(M\otimes A^1[\mathfrak g, \mathfrak g', \psi])$ is automatically a $W^\phi$-module. The idea is that there is a blockwise equivalence of categories of $V^{-\psi}$ and $W^\phi$ whose underlying functor is 
the BRST-cohomology against the kernel VOA. This functor should also provide isomorphisms on spaces of intertwining algebras, which is indeed true for $\mathfrak g = \mathfrak{gl}_1$ by \cite{Creutzig:2021bmz}. From a conformal field theory perspective one prefers a precise matching of correlation functions and similarly from the quantum geometric Langlands point of view one is interested in an equivalence of spaces of conformal blocks. 
Our aim is to provide this conformal field theory perspective. If $\mathfrak g = \mathfrak{gl}_1$, then we already succeeded \cite{Creutzig:2021cyl}. 

We now want to turn to non-abelian underlying Lie algebras. The most interesting case is the large $\mathcal N = 4$ superconformal algebra that has been realized as a BRST-cohomology of two algebras associated to the exceptional Lie superalgebra $\mathfrak{d}(2, 1; 1-\psi)$ at level one \cite{Creutzig:2019kro}. In fact 
$\mathfrak{d}(2, 1; 1-\psi)$ at level one is exactly the kernel VOA $A^1[\mathfrak{sl}_2, \mathfrak{sl}_2,  \psi])$ \cite{Creutzig:2017uxh} and the $\mathcal N = 4$ superconformal algebra is a simple current extension of $H^{\text{BRST}}_{\mathfrak{sl}_2}(A^1[\mathfrak{sl}_2, \mathfrak{sl}_2,  -\psi]) \otimes A^1[\mathfrak{sl}_2, \mathfrak{sl}_2,  \psi]))$. 

In order to lift dualities of VOAs of this type to correspondences of conformal field theory we have to understand how  to implement the BRST-cohomology \eqref{BRST-VOA} on the level of correlation functions. The first step for this is a suitable first order formulation of the conformal field theory.

\section{First order formulation of coset models}
\label{sec:1storder}

In the analysis of this paper,  we utilize the first order formulation of coset models in \cite{Creutzig:2021ykz} which refines the proposal of \cite{Gerasimov:1989mz,Kuwahara:1989xy} along with the reduction methods of \cite{Hikida:2007tq,Hikida:2008pe}.
In this section, we review  the first order formulation of cosets in \cite{Creutzig:2021ykz} using the BRST formulation 
\cite{Gawedzki:1988nj,Karabali:1988au,Karabali:1989dk,Hwang:1993nc} and the Kugo-Ogima method \cite{Kugo:1979gm}. In order to illustrate the formulation, we analyze the simplest and non-trivial example the coset \eqref{Lcoset} with $n=2$
and relate it with Liouville field theory.

In the BRST formulation, the effective action of the coset theory \eqref{Lcoset} 
is given by
\begin{align}
	S =  S_{k}^\text{WZNW} [\phi, \beta, \gamma] + S_\psi [\psi] +  S_{-k + 5}^\text{WZNW} [\tilde \phi,\tilde  \beta, \tilde \gamma]  + S_{bc} [b^a ,c_a ] \, . \label{BRSTaction}
\end{align}
For the action of $SL(2)$ WZNW model, we use the first order formulation as
\begin{align}
	S_k^\text{WZNW} [\phi , \beta , \gamma ] = \frac{1}{2 \pi} \int d^2 w 
	\left [ \partial \phi \bar \partial \phi + \beta \bar \partial \gamma + \bar \beta \partial \bar \gamma + \frac{b}{4} \sqrt{g} \mathcal{R} \phi  + \lambda \beta \bar \beta e^{2 b \phi}\right] \label{SL2action}
\end{align}
with $b = 1/\sqrt{k-2}$.
Here $g_{\mu \nu}$ represents the worldsheet metric and $g = \det g_{\mu \nu}$. Furthermore, $\mathcal{R}$ is the Ricci curvature with respect to the worldsheet metric.
We denote the  $\mathfrak{sl}(2)$ currents at level $k$ by $J^a$ with $a= \pm ,3$, which are expressed as
\begin{align}
J^+ (z) = \beta \, , \quad J^3(z) = b^{-1 } \partial \phi + \beta \gamma \, , \quad 
J^- (z)= \beta \gamma \gamma + 2 b^{-1} \gamma \partial \phi - k \partial \gamma \, .
\end{align}
 For the third action, the level $k$ is replaced by $- k+ 1 - 2 c_{SL(2)} = -k + 5$, where $c_{SL(2)}=-2$ is the dual Coxeter number of $SL(2)$. We also use $\tilde b = 1/\sqrt{- k + 3}$.  The $\mathfrak{sl}(2)$ currents at level $\tilde k = -k+5$ is denoted by $\tilde J^a$ with $a= \pm ,3$. The second action $S_\psi [\psi] $ is for complex free fermions $\psi^\pm, \bar \psi^\pm$ and given by
\begin{align}
	S_{\psi}[\psi] = \frac{1}{2 \pi} \int d^2 w  \left[  \psi^+ \bar \partial \psi^- + \bar \psi^+ \partial \bar \psi^- \right] \, .
\end{align}
The conformal weights of the fermions are $1/2$.
We frequently use its bosonized formulation  as
\begin{align}
\psi^\pm = e^{\pm i \sqrt{2}Y^L} \, , \quad Y^L (z) Y^L (0) \sim - \frac12 \ln z  \, .
\end{align}
We also introduce $Y^R(\bar z)$ in a similar manner and set $Y = Y^L + Y^R$. This model has the symmetry of $\mathfrak{sl}(2)$ current algebra with level $-1$, and the generators are 
\begin{align}
J^+ _\psi = - e^{2 i Y^L} \, , \quad J^3_\psi = i \partial Y^L \, , \quad J^-_\psi = e^{-2 i Y^L} \, .
\end{align}
The other action $ S_{bc} [b^a ,c_a ]$ is for the BRST ghosts and given by
\begin{align}
	S_{bc}[b^a , c_a] = \frac{1}{2 \pi} \int d^2 w \sum_{a = \pm , 3 } \left[  b^a \bar \partial c_a +  \bar b^a \partial \bar c_a \right] \, , \label{ghostaction}
\end{align}
where the conformal weights of $(b^a,c_a)$ are $(1,0)$. We can construct $\mathfrak{sl}(2)$ currents at level $2 c_{SL(2)} = - 4$, which are denoted by $J^{a}_{bc}$  with $a= \pm ,3$. The BRST charge for the holomorphic part is then given by
\begin{align}
	Q = \oint \frac{dz}{2 \pi i}  \left[ c_a (z) \left(J^a (z) + J^a_\psi (z) + \tilde J^a (z) + \frac{1}{2} J_{bc}^a (z) \right) \right] \, ,
\end{align}
which satisfies the nilpotency condition as $Q^2 = 0$.
The physical states are given by elements of $Q$-cohomology in the BRST formulation.

We consider the correlation functions of vertex operators of the form
\begin{align}
V = \mathcal{P} (\gamma , \tilde \gamma) e^{2 bj \phi} e^{2 i (s Y^L + \bar s Y^R)} e^{2 \tilde b \tilde \jmath \tilde \phi} \, , 
\end{align}
where $ \mathcal{P}$ is a function of $\gamma , \tilde \gamma$ (and $\bar \gamma , \bar{\tilde \gamma}$).
We define an operator 
\begin{align}
N^{\beta \gamma} = \sum_{m=-\infty}^\infty \beta_{-m} \gamma_m \label{Nbg}
\end{align}
and decompose the BRST charge as 
$Q = Q_1 + Q_0 + Q_{-1}$ by its eigenvalue. In particular, we have $Q_1 = \sum_m \beta_{-m} c_{ + , m}$, which satisfy $(Q_1)^2 = 0$ as well. Defining $R = \sum_m \gamma_{-m} b^+_m$, we introduce
\begin{align}
S = \{ Q_1 , R\} = \sum_{m=1}^\infty \beta_{-m} \gamma_m + \sum_{m=0}^\infty \gamma_{-m} \beta_m 
 + \sum_{m=1}^\infty b^+_{-m} c_{+,m} - \sum_{m=0}^\infty c_{+ , -m} b^+_m \, .
\end{align}
The terms including the BRST ghosts will be neglected in the current analysis.
Since $S$ commute with $Q_1$, we can deal with eigenstates of $S$ in $Q_1$-cohomology as $S | s \rangle = s | s \rangle$. If $s$ is non-zero, then we can rewrite the eigenstate as
\begin{align}
|s \rangle = \frac{1}{s} \{ Q_1 , R\} |s \rangle = \frac{1}{s} Q_1 R | s \rangle \, , \label{zero}
\end{align}
which is $Q_1$-exact. Thus, non-trivial elements of $Q_1$-cohomology come from the sector with zero eigenvalue.
With a new operator $U = \{ Q_0 + Q_{-1} , R \}$, a new state can be defined by 
\begin{align}
|s ' \rangle = (1 - S^{-1} U + S^{-1} U S^{-1} U - \cdots ) |s \rangle \, . \label{sp}
\end{align}
If $S |s\rangle =0 $, then we can show that $(S+ U) | s ' \rangle = 0$. This means that $|s ' \rangle $ can be an non-trivial element of $Q$-cohomology as well. In other words, a non-trivial element of $Q$-cohomology can be put in the form of \eqref{sp}.

In the effective action \eqref{BRSTaction} with \eqref{SL2action}, an interaction term includes $\beta$, which  increases the eigenvalue of $N^{\beta \gamma}$ defined in \eqref{Nbg}. The field can be removed by subtracting a BRST exact term as
\begin{align}
\beta (w) - \oint _w \frac{dz}{2 \pi i} b^+ (z) Q(w) = e^{2 i Y^L}(w) - \tilde \beta (w)  - J^+_{bc} (w)\, . \label{betar}
\end{align}
After removing $\beta$ in the action, there is no operator increasing the eigenvalue of $N^{\beta\gamma}$ defined in \eqref{Nbg}. On the other hand, we have shown that elements of $Q$-cohomology can be put into the form \eqref{sp} and the terms except for the first one in the right hand side decrease the eigenvalue of $N^{\beta\gamma}$. Therefore, we can conclude that $\beta,\gamma$ can be removed from the system once $\beta$ in the interaction term is replaced by \eqref{betar}.

We can restrict the form of vertex operators furthermore. We notice that the operators
\begin{align}
&J^{\text{tot},3}_0 \equiv \{Q , b_0^3\} = J^3_0 + J^3_{\psi , 0} + \tilde{J}_0^3 + J_{bc,0}^3 \, , \label{Jtot} \\
&L^\text{tot}_0 \equiv \{ Q , \tfrac{1}{k-3} \sum_n (J^a_n + J^a_{\psi,n} - \tilde J^a_{n} ) b_{-n ,a}  \label{Ltot}\} 
\end{align}
commute with $Q$, so we can take eigenfunctions of these operators in the $Q$-cohomology.
Moreover, from the arguments around \eqref{zero}, we can conclude that non-trivial elements of $Q$-cohomology only come from the sector with zero eigenvalues.
Since $\gamma$-dependence (and $\bar \gamma$-dependence) is removed, the vertex operators take the form
\begin{align}
V = \tilde \gamma^m { \bar{\tilde \gamma}}^{\bar m} e^{2 bj \phi} e^{2 i ( s Y^L + \bar s Y^R)} e^{2 \tilde b \tilde \jmath \tilde \phi} \, .
\end{align}
The conditions of zero eigenvalues for the operators \eqref{Jtot} and \eqref{Ltot} become%
\footnote{There is a term $+1$ in the left hand side of the first equation. This term can be explained by the definition of vacuum for the BRST ghosts, see \cite{Hwang:1993nc,Creutzig:2021ykz}.}
\begin{align}
- j - m + s - \tilde \jmath + 1 = 0 \, , \quad - \frac{(j-s)(j-s -1)}{k-3} + \frac{\tilde \jmath (\tilde \jmath -1)}{k-3} = 0
\end{align}
and that with $m,s$ replaced by $\bar m,\bar s$.
From these conditions, the vertex operators can be put into  the form
\begin{align}
V = e^{2 b j \phi} e^{2 i s Y } e^{2 \tilde b (1 - j + s) \tilde \phi} \, .
\end{align}
Since there is no $\tilde \gamma$-dependence here, we can neglect $\tilde \beta$ in the action as well.

We may rotate the fields as
\begin{align}
b \phi + i Y + b' \phi ' \, , \quad  - i \phi + b Y = b' Y' \, , \quad b' = \sqrt{\frac{3-k}{k-2}} \, , \label{Lrotation}
\end{align}
then the vertex operators become
\begin{align}
V = e^{2 ( (b' + 1/b') j - 1/b' s) \phi ' - 2 i \tilde b (j-s) Y' + 2 \tilde b (1 - j + s) \tilde \phi} \, .
\end{align}
The background charges for $Y' , \tilde \phi$ are $Q_{Y'} = - i \tilde b , Q_{\tilde \phi} = \tilde b$, respectively, and moreover there is no interaction terms for $Y' , \tilde \phi$. Therefore, after performing reflection relations, contributions from $Y' , \tilde \phi$  to the correlation functions cancel out with each other.
In summary, the $N$-point function of primary operators in the coset \eqref{Lcoset}  can be evaluated as
\begin{align}
\left \langle \prod_{\nu=1}^N V_\nu (z_\nu) \right \rangle  \, , \quad V_\nu (z_\nu)= e^{2 ( (b' + 1/b') j_\nu - 1/b' s_\nu) \phi '   (z_\nu)}
\end{align}
with the action
\begin{align}
S = \frac{1}{2 \pi} \int d^2 w \left[ \partial \phi ' \partial \phi ' + \frac{\sqrt{g} \mathcal{R}}{4} (b' + 1/b') \phi ' + \lambda e^{2 b' \phi '} \right] \, .
\end{align}
This is nothing but an $N$-point function of Liouville field theory.

Here we compare the analysis to that of \cite{Gerasimov:1989mz,Kuwahara:1989xy}. As they proposed, $\beta,\gamma$ from the numerator are canceled with $\tilde \beta , \tilde \gamma$ from the denominator. Moreover, the free boson corresponding to Cartan direction is taken to be orthogonal to $\tilde \phi$ from the denominator via the rotation \eqref{Lrotation}. However, from their analysis, it is not clear how the interaction term of Liouville field theory arises. In our formulation, this comes from the replacement \eqref{betar}, which may be the most important point in our approach. In the succeeding sections, we apply this analysis to more involved examples.

\section{FZZ-triality}
\label{sec:coset1}

Applying the method reviewed in the previous section,
we examine  correlation functions of primary operators in the coset \eqref{trialitycoset}.
We derive correlator correspondences between the coset \eqref{trialitycoset} and sine-Liouville theory in the next subsection and those between the cosets \eqref{trialitycoset} and \eqref{cigar} in subsection \ref{sec:coset2}. For the two cases, we use different first order formulations of $SL(2|1)_k$ in the numerator of \eqref{trialitycoset}. In subsection \ref{sec:fdual}, we extend the analysis by introducing $\mathcal{N}=2$ superconformal symmetry.

\subsection{A realization of the coset}

\label{sec:freecoset}

In order to apply our method to the coset \eqref{trialitycoset}, we first need  a  first order formulation of $SL(2|1)$ WZNW model. In this subsection, we use the action obtained in  \cite{Hikida:2007sz,Creutzig:2011qm}.
We express the element of  Lie supergroup $SL(2|1)$ as 
\begin{align}
g = g_{-1} g_{-\frac{1}{2}} g_0 g_{\frac{1}{2}} g_{1} \label{para1}
\end{align}
with
\begin{align}
g_{-1} = e^{\gamma E^+} \, , \quad 
g_{-\frac{1}{2}} = e^{\theta_1 F^+ +  \theta_2 G^+} \, , \quad 
g_{0} = e^{2 \phi_1 H + 2 \phi_2 I} \, , \quad 
g_{\frac{1}{2}} = e^{\bar \theta_2 F^- + \bar \theta_1 G^-} \, , \quad 
g_{1} = e^{\bar \gamma E^-} \, .
\end{align}
We use the notation for generators of $\mathfrak{sl}(2|1)$ given in appendix \ref{sec:convention}.
The grading is made by $H$.
With the help of Polyakov-Wiegmann identity, we then obtain
\begin{align}
\begin{aligned}
S = \frac{k}{2 \pi} \int d^2 z \left[ \bar \partial \phi _1 \partial \phi _1  - \bar \partial \phi _2  \partial \phi _2   - e^{ - \phi _1 - \phi _2 } \bar \partial \theta_1  \partial \bar \theta_1 + e^{ - \phi _ 1  + \phi _ 2 } \bar \partial \theta_2 \partial \bar \theta_2   \right. \\
  +   \left. e^{-2 \phi _ 1 } \left(\partial \bar \gamma - \tfrac12 (  \bar \theta_2 \partial \bar \theta_1 +  \bar \theta_1 \partial \bar \theta_2 )\right) \left(\bar \partial \gamma - \tfrac12 (  \theta_2 \bar \partial \theta_1 + \theta_1 \bar \partial \theta_2  ) \right) \right ] \, .
\end{aligned}
\end{align}
Introducing auxiliary fields, we can rewrite the action as
\begin{align}
	\begin{aligned}
		S &= \frac{1}{2 \pi} \int d^2 z \left[ k \bar \partial \phi _1 \partial \phi _1  - k \bar \partial \phi _2 \partial \phi _2 + \beta \bar \partial \gamma + \bar \beta \partial \bar \gamma + \sum_{a=1}^2 ( p_a \bar \partial \theta_a  +  \bar p_a \partial \bar \theta_a )   -  \frac{1}{k} e^{ 2 \phi _ 1 } \beta \bar \beta  
		\right] \\
		&\quad +  \frac{1}{2 k \pi} \int d^2 z \left[  e^{\phi _1 + \phi _ 2 } (p_1 + \tfrac12 \beta \theta_2 ) (\bar p_1 + \tfrac12 \bar \beta \bar \theta_2 ) - e^{\phi _ 1 - \phi _ 2} (p_2 + \tfrac12 \beta \theta_1 ) (\bar p_2 + \tfrac12 \bar \beta \bar \theta_1 ) \right] \, .
	\end{aligned} \label{1stcla}
\end{align}
The equations of motion lead to
\begin{align}
	\begin{aligned}
&	\beta = k e^{ - 2 \phi_1} (\partial \bar \gamma - \tfrac12 (\bar \theta_2 \partial \bar \theta_1 + \bar \theta_1 \partial \bar \theta_2)) \, ,  \\
&	p_1 + \tfrac12 \beta \theta_2 =  k e^{- \phi_1 - \phi_2 } \partial \bar \theta_1  \, , \quad
	p_2 + \tfrac12 \beta \theta_1 = - k e^{- \phi_1 + \phi_2 } \partial \bar \theta_2 \, , \\
&\bar 	\beta = k e^{ - 2 \phi_1} (\bar \partial \gamma - \tfrac12 ( \theta_2 \bar \partial  \theta_1 +  \theta_1 \bar \partial  \theta_2)) \, ,  \\
&	\bar p_1 + \tfrac12 \bar \beta \bar \theta_2 = - k e^{- \phi_1 - \phi_2 } \bar \partial \theta_1  \, , \quad
	\bar p_2 + \tfrac12 \bar \beta \bar \theta_1 =  k e^{- \phi_1 + \phi_2 } \bar \partial \theta_2 \, . 
	\end{aligned} \label{eom}
\end{align}
Taking into account quantum corrections due to the change of variables (see, e.g., (2.8) and (2.9) of \cite{Hikida:2007sz}), we have
\begin{align}
\begin{aligned}
S &= \frac{1}{2 \pi} \int d^2 z \left[ \bar \partial \phi _1 \partial \phi _1  - \bar \partial \phi _2 \partial \phi _2 + \beta \bar \partial \gamma + \bar \beta \partial \bar \gamma + \sum_{a=1}^2 ( p_a \bar \partial \theta_a  +  \bar p_a \partial \bar \theta_a )   -  \frac{1}{k} e^{ 2 b\phi _ 1 } \beta \bar \beta  
\right] \\
&\quad + \frac{1}{2 k \pi} \int d^2 z \left[  e^{b (\phi _1 + \phi _ 2 )} (p_1 + \tfrac12 \beta \theta_2 ) (\bar p_1 + \tfrac12 \bar \beta \bar \theta_2 ) - e^{b (\phi _ 1 - \phi _ 2)} (p_2 + \tfrac12 \beta \theta_1 ) (\bar p_2 + \tfrac12 \bar \beta \bar \theta_1 ) \right] 
\end{aligned} \label{sl21action1}
\end{align}
with $b = 1/\sqrt{k-1}$. 
Since the interaction term with $ e^{ 2 b\phi} \beta \bar \beta $ can be induced from the rest two, we simply neglect the term.

The symmetry of $SL(2|1)$ WZNW model is the $\mathfrak{sl}(2|1)$ current algebra.
In the first order formulation, the generators are written in terms of fields and commute with interaction terms.
We find that they are given by
\begin{align}
&F^+ =  p_1 - \frac12 \beta \theta_2 \, , \quad G^+ = -  p_2 +  \frac12 \beta \theta_1 \, , \quad E^+ =  \beta  \, ,   \nonumber  \\
&H = b^{-1} \partial \phi_1 + \gamma \beta + \frac12 (p_1 \theta_1 + p_2 \theta_2) \, ,\quad
I = - b^{-1} \partial \phi_2 + \frac12 (p_1 \theta_1 - p_2 \theta_2) \, ,  \nonumber  \\
&F^- =  - b^{-1} \partial \phi_1 \theta_2 + b^{-1} \partial \phi_2 \theta_2  - \frac12 \gamma \beta \theta_2 - \frac12 p_1 \theta_1 \theta_2 -  \left(\frac12-k\right) \partial \theta_2 + \gamma p_1 \, , \\
&G^- =  b^{-1} \partial \phi_1 \theta_1 + b^{-1} \partial \phi_2 \theta_1 + \frac12 \gamma \beta \theta_1 - \frac12  p_2 \theta_1 \theta_2 + \left(  \frac12 - k \right) \partial \theta_1  - \gamma p_2 \, ,  \nonumber  \\
&E^- = -  \gamma \gamma \beta -  \gamma p_1 \theta_1 -  \gamma p_2 \theta_2 -  2 b^{-1} \partial \phi_1 \gamma + b^{-1} \partial \phi_2 \theta_1 \theta_2 -  \frac12  (1-k)  ( \theta_1 \partial \theta_2 -  \partial  \theta_1 \theta_2 )   + k \partial \gamma  \, . \nonumber 
\end{align} 
Here the normal ordering prescription is assumed for the products of fields.
It will be convenient to bosonize the fermionic systems $(p_a , \theta_a)$ as
\begin{align}
p_a = e^{i  X_a^L } \, , \quad  \theta_a = e^{ - i  X_a^L} \, , \quad 
X_a^L (z) X_b^L (0) \sim - \delta_{a,b}\ln z  \, .\label{ptbos} 
\end{align}
We also define $X^R_a$ analogously from $(\bar p_a , \bar \theta_a )$ and consider linear combinations
$X_a = X_a^L + X_a^R$.
With these bosons, the Cartan generators are expressed as
\begin{align}
H = b^{-1} \partial \phi_1 + \gamma \beta + \frac{i}{2} ( \partial X_1 + \partial X_2) \, ,\quad
I = - b^{-1} \partial \phi_2 + \frac{i}{2} (\partial X_1 - \partial X_2  )
\end{align}
in particular.

Now that we have the first order formulation of $SL(2|1)$ WZNW model, we can apply the procedure of \cite{Creutzig:2021ykz} reviewed in the previous section. Firstly, two new bosons $\hat \phi_1 , \hat \phi_2$ are introduced by
\begin{align}
H = \sqrt{k - 2} \partial \hat \phi_1 + \beta \gamma \, , \quad I = \sqrt{k} \partial \hat \phi_2 \, .
\end{align}
Then the field space of the coset model \eqref{trialitycoset} is spanned by the fields $\phi_1 , \phi_2 , X_1 , X_2$ but orthogonal to the new bosons $\hat \phi_1 , \hat \phi_2$. The $(\beta,\gamma)$-system for the free filed realization of $\mathfrak{sl}(2|1)$ are removed as explained above. 
In other words, the orthogonal space can be generated by $\phi , \chi$, where we have defined as
\begin{align}
\sqrt{k-2} \phi = - i \phi_1  + \frac{1}{2 b} (X_1 + X_2) \, , \quad
\sqrt{k}  \chi = - i \phi_2  + \frac{1}{2 b} (X_1 - X_2) \, .\label{Xpm1} 
\end{align}
With the new variables, the correlation functions are written as
\begin{align}
\left \langle \prod_{\nu=1}^N \Psi_\nu (z_\nu) \right \rangle \, , \quad
 \Psi_\nu (z_\nu) = e^{2 \hat b j \phi + i \frac{2}{ \sqrt{\kappa}} (m \chi_L + \bar m \chi_R)}  \label{corr}
\end{align}
with
\begin{align}
 \kappa = \frac{k}{k-1} \, .  \label{kappa}
\end{align}
The correlation function is evaluated with the action%
\footnote{The coefficients in front of interaction terms are modified by shifting fields. The same procedure will be performed in later analysis as well.}
\begin{align}
\begin{aligned}
S &= \frac{1}{2 \pi} \int d^2 z \left[ \bar \partial \phi \partial \phi +  \bar \partial \chi \partial \chi 
+ \frac{\hat b}{4} \sqrt{g} \mathcal{R} \phi  + 2 \lambda e^{ \phi / \hat b} \cos \left( \sqrt{\kappa} \chi \right)   \right] \, , 
\end{aligned}  \label{sineaction}
\end{align}
where we set $\hat b = 1/\sqrt{\kappa -2}$.  The background charge for $\phi$ arises from those for $X_1$ and $X_2$. 
This is nothing but the action of sine-Liouville theory.
In this way, we can write down the $N$-point function of the coset \eqref{trialitycoset} in terms of sine-Liouville theory as a direct consequence of the first order formulation reviewed in the previous section.

\subsection{Another realization of the coset}

\label{sec:coset2}

In the previous subsection, we have shown that correlation functions of the coset \eqref{trialitycoset} directly reduce to those of sine-Liouville theory. Combined with the FZZ-duality proven in \cite{Hikida:2008pe}, we can  derive the correlator correspondences between the two cosets  \eqref{trialitycoset}  and \eqref{cigar}. Instead of doing so, we directly relate the correlation functions of these two cosets by choosing a different first order formulation of $SL(2|1)$ WZNW model in this subsection.

It is known that there are several different free field realizations of the same affine Lie algebra. In terms of WZNW model, this corresponds to the fact that there are several ways to parameterize group elements.
In \cite{Hikida:2007sz,Creutzig:2011qm} or \eqref{para1}, we use the grading corresponding to the Cartan subalgebra generated by $H$. This is a natural choice, but here we adopt a different one with the grading of linear combination $H + I$. Then the element of Lie supergroup $SL(2|1)$ can be expressed as 
\begin{align}
	g = g_{+} g_0 g_{-} 
\end{align}
with
\begin{align}
	g_{+} = e^{\gamma E^+} e^{\theta_1 F^+} \, , \quad 
	g_{0} = e^{ \theta_2 F^-} e^{2 \phi_1 H + 2 \phi_2 I} e^{\bar \theta_2 G^+} \, , \quad 
	g_{-} = e^{\bar \theta_1 G^-}e^{\bar \gamma E^-} \, .
\end{align}
Note that $g_+$ ($g_-$) is put on the left (right) side of $g$ but a factor including $G^+$ ($F^-$) is put on the right (left) side inside $g_0$. Otherwise, the expression of Lie superalgebra becomes identical to the one in the previous subsection.
Using the Polyakov-Wiegmann identity again, we obtain
\begin{align}
	\begin{aligned}
		S &= \frac{k}{2 \pi} \int d^2 z \left[ \bar \partial \phi _1 \partial \phi _1  - \bar \partial \phi _2  \partial \phi _2  + e^{-2 \phi _ 1 }  \partial \bar \gamma  \bar \partial \gamma  \right. \\
		& \quad - \left. e^{ - \phi _1 - \phi _2 } (\bar \partial \theta_1 + \theta_2 \bar \partial \gamma )(  \partial \bar \theta_1 + \bar \theta_2 \partial \bar \gamma ) - e^{  \phi _ 1  -  \phi _ 2 }  \bar \partial \theta_2 \partial \bar \theta_2  \right ] \, .
	\end{aligned}
\end{align}
Introducing auxiliary fields, we rewrite the action as
\begin{align} \label{WZNW2}
		S &= \frac{1}{2 \pi} \int d^2 z \left[ \bar \partial \phi _1 \partial \phi _1  - \bar \partial \phi _2 \partial \phi _2 
		+ \frac{b}{4} \sqrt{g} \mathcal{R} (\phi_1 - \phi_2) 
		+ \beta \bar \partial \gamma + \bar \beta \partial \bar \gamma + \sum_{a=1}^2 ( p_a \bar \partial \theta_a  +  \bar p_a \partial \bar \theta_a ) 
		\right] \nonumber \\
		&\quad - \frac{1}{2 k \pi} \int d^2 z \left[ e^{ 2 b\phi _ 1 } (\beta - p_1 \theta_2 ) ( \bar \beta - \bar p_1 \bar \theta_2 )  - e^{b ( \phi _1 + \phi _ 2 )} p_1 \bar p_1  - e^{b (- \phi _ 1 + \phi _ 2)} p_2 \bar p_2 \right] 
\end{align}
with $b = 1/\sqrt{k-1}$ as above. Here we have already included quantum corrections.
We will neglect the interaction term $e^{b (\phi_1 + \phi_2)} p_1 \bar p_1$ since it can be generated by the rest two.

As mentioned above, a different parameterization of group element in WZNW model corresponds to a different free field realization of the symmetry algebra. The corresponding free field realization of affine Lie superalgebra $\mathfrak{sl}(2|1)$ is given by fields in the kinetic terms and generators made of the free fields should commute with the interaction terms.
We find these generators as
\begin{align}
	\begin{aligned}
		&F^+ = p_1 \, , \quad G^+ = \beta \theta_1 - b^{-1} \partial \phi_1 \theta_2 + b^{-1} \partial \phi_2 \theta_2 + (1 -k) \partial \theta_2  \, , \quad E^+ = \beta  \, , \\
		&H = b^{-1} \partial \phi_1 + \gamma \beta + \frac12 (p_1 \theta_1 - p_2 \theta_2) \, ,\quad
		I = - b^{-1} \partial \phi_2 + \frac12 (p_1 \theta_1 + p_2 \theta_2) \, , \\
		&F^- = - p_2 + \gamma p_1 \, , \quad
		G^- = b^{-1} \partial \phi_1 \theta_1 + b^{-1} \partial \phi_2 \theta_1 + \gamma \beta \theta_1  + p_2 \theta_1 \theta_2 - k  \partial \theta_1  \\
		& \qquad \qquad  \qquad \qquad  \qquad \qquad  - b^{-1} \partial \phi_1 \gamma \theta_2 + b^{-1} \partial \phi_2 \gamma \theta_2 - (k-1) \gamma \partial \theta_2   \, , \\
		&E^- = - \gamma \gamma \beta - \gamma p_1 \theta_1 + \gamma p_2 \theta_2- 2 b^{-1} \partial \phi_1 \gamma + p_2 \theta_1 + k \partial \gamma  \, .
	\end{aligned}
\end{align} 
The Cartan generators can be expressed as
\begin{align}
	H = b^{-1} \partial \phi_1 + \gamma \beta + \frac{i}{2} ( \partial X_1 - \partial X_2) \, ,\quad
	I = - b^{-1} \partial \phi_2 + \frac{i}{2} (\partial X_1 + \partial X_2  )
\end{align}
in terms of $X_a$ introduced in \eqref{ptbos}.

Let us apply the first order formulation reviewed in the previous section to this case.
As before we define new bosons $\hat \phi_1 , \hat \phi_2$ by
\begin{align}
	H = \sqrt{k - 2} \partial \hat \phi_1 + \beta \gamma \, , \quad I = \sqrt{k} \partial \hat \phi_2 \, .
\end{align}
The field space for the numerator is generated by $\phi_1 ,  \phi_2 , X_1  , X_2$ if $\beta , \gamma$ are neglected, 
and the field space orthogonal to $\hat \phi_1 , \hat \phi_2$ may be spanned by $\phi, \chi$ with
\begin{align}
	\sqrt{k-2} \phi= - i \phi_1  + \frac{1}{2 b} (X_1 - X_2) \, , \quad
	 \sqrt{k} \chi =  -i \phi_2  + \frac{1}{2 b} (X_1 + X_2) \, . \label{Xpm2}
\end{align}
The correlation functions are of the form as \eqref{corr} but now the action is
\begin{align}
	\begin{aligned}
		S &= \frac{1}{2 \pi} \int d^2 z \left[ \bar \partial \phi \partial \phi +  \bar \partial \chi \partial \chi 
		+ \frac{1}{4} \sqrt{g} \mathcal{R}(Q_\phi  \phi + Q_\chi \chi ) - \frac{1}{k} (  e^{ 2 \phi / \hat b} - e^{- \phi / \hat b + i \sqrt{\kappa } \chi }   )  \right] \, .
	\end{aligned}
\end{align}
Here we set
\begin{align}
Q_\phi  = \hat b + \frac{1}{\hat b} \, , \quad Q_\chi = - i \sqrt{\kappa} 
\end{align}
with $ \kappa$ given in \eqref{kappa} and $\hat b = 1/\sqrt{\kappa -2}$.   The action coincides with (3.14) of \cite{Hikida:2008pe}, which directly arises from the reduction of $SL(2) \otimes U(1) \, (\otimes \text{BRST ghosts})$ description for the coset $SL(2)/U(1)$.
The first term of the interaction comes from (the dual of) the interaction term in the first order formulation of $SL(2)$ WZNW model and the second one stems from the extra insertions of degenerate operators in the Ribault-Teschner relation \cite{Ribault:2005wp,Ribault:2005ms,Hikida:2007tq}.

\subsection{Fermionic dualities}
\label{sec:fdual}

In this subsection, we extend the analysis by introducing $\mathcal{N}=2$ supersymmetry.
The fermionic version of the coset is given by
\begin{align}
\frac{ SL(2|1)_k \otimes SO(2)_1}{ SL(2)_k \otimes U(1) } \, . \label{scoset}
\end{align}
The factor $SO(2)_1$ can be generated by a complex fermion $\psi^\pm$ with conformal weight $1/2$,
and the bosonization formula
\begin{align}
	\psi^\pm = e^{\pm i Y^L } \, , \quad 
	Y^L (z) Y^L (0) \sim - \ln z  \label{bc}
\end{align}
is used.
We define $Y^R$ in a similar way and introduce $Y = Y^L + Y^R$.
Notice that the Cartan directions of the denominator algebra are generated by
\begin{align}
  H \, , \quad  I + \tfrac12 \psi^+\psi^- = I + \frac{i}{2} \partial Y \, .\label{superden}
\end{align}

We first use the free field realization of  affine Lie superalgebra $\mathfrak{sl}(2|1)$ in subsection \ref{sec:freecoset}.
In this case, the orthogonal space is generated by $\hat Y$ as well as $\phi,\chi$ defined in \eqref{Xpm1}.
Here $\hat Y$ may be given by
\begin{align}
 \sqrt{\frac{k  (1 + 2 k )}{2}} \hat Y =  i b^{-1} \phi_2 + \frac{1}{2} (X_1 - X_2) -  k Y \, .
\end{align}
We further rotate the fields as
\begin{align}
 	\sqrt{\frac{k}{k-1}} \chi \to  \sqrt{\frac{2 -k}{k-1}} \chi+   \hat Y  \,  , \quad
	 \sqrt{\frac{k}{k-1}} \hat Y \to  - 2 \chi + \sqrt{\frac{2 -k}{k-1}}  \hat Y \, . \label{rotation}
\end{align}
The correlation functions are of the form as 
\begin{align}
\left \langle \prod_{\nu=1}^N \Psi_\nu (z_\nu) \right \rangle \, , \quad
 \Psi_\nu (z_\nu) = e^{2 \hat b j \phi + i (s Y^L + \bar s Y^R) + i \frac{2}{ \sqrt{\kappa}} (m \chi_L + \bar m \chi_R)}  \label{scorr}
\end{align}
and the action is 
\begin{align}
	\begin{aligned}
		S &= \frac{1}{2 \pi} \int d^2 z \left[ \bar \partial \phi \partial \phi +  \bar \partial \chi \partial \chi + \psi_+ \bar  \partial \psi_- + \bar \psi_+ \partial \bar \psi_-
		+ \frac{\hat b}{4} \sqrt{g} \mathcal{R} \phi  \right]  \\
	& \quad 	+  \frac{\lambda}{2 \pi}  \int d^2 z \left[   \psi_+ \bar \psi_+ e^{ \hat b {}^{-1} (\phi + i \chi )} 
		+ \psi_- \bar \psi_- e^{ \hat b {}^{-1} (\phi - i \chi )} \right] \, .
	\end{aligned} \label{sLaction}
\end{align}
Here we have introduced new complex fermions $\psi^\pm$ using \eqref{bc} but with $Y$ replaced by $\hat Y$.
This is nothing but the action of $\mathcal{N}=2$ super Liouville theory.
In this way, we have shown that $N$-point functions of the super coset \eqref{scoset} can be reduced to those of  $\mathcal{N}=2$ super Liouville theory.

We next use  another free field realization of  affine Lie superalgebra $\mathfrak{sl}(2|1)$ in subsection \ref{sec:coset2}.
The field space orthogonal to the denominator algebra with Cartan directions \eqref{superden} is generated by $\phi , \chi$ in \eqref{Xpm2} and $\hat Y$ defined via
\begin{align}
 \sqrt{\frac{ k(1 + 2 k )}{2}} \hat Y = i b^{-1} \phi_2 + \frac{1}{2} (X_1 + X_2) - k Y \, .
\end{align}
We also perform the rotation of fields in \eqref{rotation}.
The correlation functions are of the form as \eqref{scorr} and
the action is now
\begin{align}
	\begin{aligned}
		S &= \frac{1}{2 \pi} \int d^2 z \left[ \bar \partial \phi \partial \phi +  \bar \partial \chi \partial \chi + \frac12 \bar \partial \hat Y \partial \hat Y 
		+ \frac{1}{4} \sqrt{g} \mathcal{R} \left( Q_\phi \phi + Q_\chi \chi + Q_Y \hat Y  \right) \right]  \\
		& \quad 	- \frac{ 1}{2 \pi k} \int d^2 z \left[   e^{  2 \hat b {}^{-1} \phi } 
		-  e^{- \hat b {}^{-1} (\phi - i \chi ) + i \hat Y  } \right] \, .
	\end{aligned}
\end{align}
The background charges are
\begin{align}
	Q_\phi = \hat b + \hat b^{-1} \, , \quad Q_\chi = - i \hat b \, , \quad Q_Y = - i \, .
\end{align}
This kinetic terms of the action are the same as those in (3.11) and the interaction terms are (3.12) in \cite{Creutzig:2010bt}.%
\footnote{The notation of $Y$ is different from  $\tilde h$ in \cite{Creutzig:2010bt} by factor $-\sqrt{2}$.}
Applying the analysis of the paper, we can thus map the correlation functions of \eqref{scoset} to those of the super cigar model described by 
\begin{align}
\frac{SL(2)_\kappa \otimes SO(2)_1}{U(1)} \, . \label{scigar}
\end{align}
In the BRST formulation, the coset can be described by $SL(2)_\kappa \otimes SO(2) _1 \otimes U(1) \,  (\otimes \text{BRST ghosts})$.
The first term of the interaction directly comes from bosonic $SL(2)_\kappa$ WZNW model and the second one stems from an interpretation of extra insertions of degenerate operators in the Ribault-Teschner relation as in the bosonic case.

\section{Theory with a $\mathfrak{d}(2,1 ; - \psi)$-structure from a coset}
\label{sec:HD}

A main purpose of this paper is to express the large $\mathcal{N}=4$ super Liouville theory in terms of a coset model \eqref{cosetY}.
Before going to it, we would like to consider a related but different problem.
Namely, we study $N$-point functions of the coset \eqref{FScoset}
and reduce them to those of a theory with a $\mathfrak{d}(2,1 ; - \psi)$-structure. The relation between the two models was examined in \cite{Feigin:2001yq}.

In the examples analyzed in the previous section, we actually needed only the first order formulation of the coset proposed in  \cite{Gerasimov:1989mz,Kuwahara:1989xy}.
However, in the example with coset \eqref{FScoset}, we have to use the first order formulation elaborated in \cite{Creutzig:2021ykz} and also the reduction methods utilized in the proof of  original FZZ-duality in \cite{Hikida:2008pe}.
We begin by describing the coset theory \eqref{FScoset} in the BRST formulation \cite{Gawedzki:1988nj,Karabali:1988au,Karabali:1989dk,Hwang:1993nc}.
The effective action is given by
\begin{align}
	S = \sum_{i=1}^3 S_{k_i}^\text{WZNW} [\phi_i, \beta_i , \gamma_i ] + S_{bc} [b^a ,c_a ] \label{FSBRSTaction}
\end{align}
with $k_3 = - k_1 - k_2  - 2 c_{SL(2)} = - k_1 - k_2  + 4$. 
For the action of $SL(2)$ WZNW model, we use the first order formulation given in \eqref{SL2action}.
We denote the  $\mathfrak{sl}(2)$ currents at level $k_i$ by $J^a_i$ with $a= \pm ,3$.
In particular, we have $J^+_i = \beta_i$. 
The other action $ S_{bc} [b^a ,c_a ]$ is for the BRST ghosts given in \eqref{ghostaction} and 
$\mathfrak{sl}(2)$ currents at level $2c_{SL(2)} = - 4$ are denoted by $J^{a}_{bc}$ with $a= \pm ,3$ as before.
The BRST charge is 
\begin{align}
	Q = \oint \frac{dz}{2 \pi i}  \left[ c_a (z) \left(\sum_{i=1}^3 J^a_i (z) + \frac{1}{2} J_{bc}^a (z) \right) \right] \, ,
\end{align}
and physical states are given by elements of the $Q$-cohomology in the BRST formulation.

As explained in section \ref{sec:1storder}, we can remove one of pairs of $(\beta_i , \gamma_i)$, say,  $(\beta_1 , \gamma_1)$. Namely,  we replace $\beta_1$ by
\begin{align}
	\beta_1 (w) - \oint \frac{dz}{2 \pi i} b^+ (z)Q (w) = - \beta_2 (w) - \beta_3 (w) - J_{bc}^+ (w)\, . \label{beta1r}
\end{align}
Furthermore, we decompose the BRST-charge by $N^{\beta_1 \gamma_1} = \sum_m \beta_{1,-m} \gamma_{1,m}$ and pick up non-trivial elements of $Q_1$-cohomology.

In this way, we can put the vertex operators of the form
\begin{align}
	V = \gamma_2^{m_2} \bar \gamma_2^{\bar m_2}\gamma_3^{m_3} \bar \gamma_3^{\bar m_3}  e^{2 ( b_1 j_1  \phi_1 + b_2 j_2 \phi_2 + b_3 j_3 \phi_3 )} 
\end{align}
with $b_i = 1/\sqrt{k_i-2}$. The conditions corresponding to zero eigenvalues of \eqref{Jtot} and \eqref{Ltot}  are now given by
\begin{align}
	\begin{aligned}
	& - j_1 - j_2 - m_2  - j_3 - m_3 + 1 = 0 \, , \\
	& - \frac{(j_1 + j_2 + m)(j_1 + j_2 + m  -1)}{k_1 + k_2 -2 } + \frac{ j_3 ( j_3 - 1)}{k_1 + k_2 - 2} = 0 
\end{aligned}\label{FScond}
\end{align}
with some integer $m$.
There is also a similar condition with $ m_2 ,  m_3 , m$ replaced by $\bar m_2 , \bar m_3 , \bar m$.
We have shown that a map can be constructed between elements of $Q_1$-cohomology and $Q$-cohomology for our restricted form of vertex operators. Thus, we can pick up an element of $Q_1$-cohomology with
our preferred choice of $m_2,m_3,m$ as long as they satisfy \eqref{FScond}.
Our choice here is
\begin{align}
 m_2 = 	- j_1 -  j_2 - j_3 +1 \, , \quad m_3 = 0 \, , \quad m = m_2 \, .
\end{align}
It is convenient to perform a reflection relation to $\phi_3$-direction as
\begin{align}
V = \gamma_2^{m_2}  \bar \gamma_2^{\bar m_2}    e^{2 ( b_1 j_1  \phi_1 + b_2 j_2 \phi_2 + b_3 (1 -j_3) \phi_3 )} 
\end{align}
with
\begin{align}
	m_2 = \bar m_2 = - j_1 - j_2 - j_3 +1 \, .
\end{align}
This choice is useful since vertex operators do not depend on $\gamma_3$ so we can neglect $\beta_3$ in the action as well.

The action and vertex operators still depend on $(\beta_2 , \gamma_2)$-system.
We deal with them by applying the reduction method of \cite{Hikida:2007tq,Hikida:2008pe}.
We rewrite the vertex operators as 
\begin{align}
	&\Psi_\nu (z_\nu) =  \int \frac{d ^2 \mu_\nu}{|\mu_\nu|^2 }V_\nu (z_\nu ) \, , \label{mbasis} \\
	&V_\nu (z_\nu) =  |\mu_\nu|^{2 j_1 + 2 j_2 + 2 j_3 - 2} e^{\mu_\nu \gamma_2 - \bar \mu_\nu \bar \gamma_2 }  e^{2 ( b_1 j^\nu_1  \phi_1 + b_2 j^\nu_2 \phi_2 + b_3 (1 -j^\nu_3) \phi_3 )} \, . \label{mbasisp} 
\end{align}
We may insert an identity operator in the coset theory as in, e.g., \cite{Hikida:2008pe,Creutzig:2021ykz}.
Now we perform $s$ spectral flow operations to $SL(2)_{k_1},SL(2)_{k_2}$ in the numerator of \eqref{FScoset} and the operation is undone in the coset theory by performing $s$ spectral flow to $SL(2)_{k_1 + k_2}$ in the denominator. In the current first order formulation, the identity operator is expressed as
\begin{align}
\mathbbm{1} = v^{(s)} (\xi) e^{s (\phi_1 /b_1 -  \phi_3 /b_3)} (\xi)\, , \label{bosonicid}
\end{align}
where the spectral flow operator $v^{(s)} (\xi)$ restricts the domain of integral over $\beta_2$ such as to have a zero of order $s$ and effectively inserts $e^{s \phi_2/b_2}$ at $w = \xi$.
We then compute the correlation function of the coset model \eqref{FScoset},
\begin{align}
\begin{aligned}
&	\left \langle  v^{(s)} (\xi) e^{s (\phi_1 /b_1 -  \phi_3 /b_3)} (\xi) \prod_{\nu=1}^N V_\nu (z_\nu) \right \rangle  \\
& \quad 	= \int _s \left [  \prod_{i=1}^3  \mathcal{D}  \phi_i  \right ]\mathcal{D} \gamma_2 \mathcal{D} \beta _2  e^{- S}\prod_{\nu=1}^N V_\nu (z_\nu)  e^{s (\phi_1 /b_1 + \phi_2 /b_2-  \phi_3 /b_3)} (\xi) \, .
\end{aligned}
\end{align}
Here the effective action is
\begin{align}
	S = \frac{1}{ 2 \pi} \int d^2 w \left[ \sum_{i=1}^3 \partial \phi_i  \bar \partial \phi_i  + \beta_2 \bar \partial \gamma_2 + \bar \beta_2 \partial \bar \gamma_2  + \frac{\sqrt{g} \mathcal{R} }{4} \left( \sum_{i=1}^3 b_i \phi_i \right) 
	+ \lambda \beta_2 \bar \beta_2 \left (  e^{2 b_1 \phi_1 } + e^{2 b_2\phi_2 } \right )  \right] \, .
\end{align}
Notice that $\beta_1$ in the interaction term is replaced by \eqref{beta1r} and the terms except for $\beta_2$ are neglected.
The subscript $s$ in the integral symbol represents the restriction of integral domain for $\beta_2$.
Integrating $\gamma_2$ out, we obtain delta functions for $\mu_\nu$ and a delta functional for $\beta_2$ as explained in \cite{Hikida:2007tq,Hikida:2008pe}.
After further integrating $\beta_2$ out, it is replaced by a function as
\begin{align}
-	\beta_2 (w) =  \sum_{\nu=1}^N \frac{\mu_\nu}{w - z_\nu} = u \frac{(w - \xi)^s \prod_{i=1}^{N-2 - s } (w - y_i)}{\prod_{\nu=1}^N (w - z_\nu)} \, = u \mathcal{B} (w ; z_\nu, y_i) \, .
\end{align}
The second equality defines a map of variables from $\mu_\nu$ to $y_i$, which  is possible only when the conditions coming from the delta functions,
\begin{align}
\sum_{\nu=1}^N \frac{\mu_\nu}{(w - \xi)^a} = 0
\end{align}
are satisfied for $a = 0,1,\ldots, s$. For more details, see, e.g., \cite{Hikida:2008pe}.

In order to remove the functions in the interaction terms, we shift $\phi_i$ as
\begin{align}
	\phi_1 + \frac{1}{2 b_1} \ln | u \mathcal{B}| ^2  \to \phi_1  \, , \quad 	\phi_2 + \frac{1}{2 b_2} \ln | u \mathcal{B}| ^2 \to \phi_2  \, , \quad 	\phi_3 - \frac{1}{2 b_3} \ln | u \mathcal{B}| ^2 \to \phi_3  \, .
\end{align}
The third one is chosen such that there would be no extra factor with $\mu_\nu$ in the vertex operator of the form \eqref{mbasisp}.
Now the correlation function becomes
\begin{align}
	\left \langle \prod_{\nu=1}^N V_\nu (z_\nu) \right \rangle 
	= 	\left \langle \prod_{\nu=1}^N \tilde  V_\nu (z_\nu)\prod_{i=1}^{N-2-s} \tilde V_b (y_i) \right \rangle  \label{corrrel}
\end{align}
with the vertex operators
\begin{align}
 &\tilde  V_\nu (z_\nu) =  e^{2 ( b_1 ( j^\nu_1 + 1/2b_1^2 ) \phi^\nu_1 + b_2( j^\nu_2 + 1/2b_2^2 ) \phi_2 + b_3 (1 -j^\nu_3 - 1/2b_3^2) \phi_3 )}  \, , \\
 &\tilde  V_b (y_i) =  e^{-  \phi_1 /b_1 -  \phi_2 /b_2  +  \phi_3 /b_3  }  \, .
\end{align}
Notice that the insertion at $w = \xi$ is canceled out. 
The right hand side is evaluated with the action 
\begin{align}
	S = \frac{1}{ 2 \pi} \int d^2 w \left[ \sum_{i=1}^3 \partial \phi_i  \bar \partial \phi_i   + \frac{\sqrt{g} \mathcal{R} }{4} \left( \sum_{i=1}^3 Q_i \phi_i  \right)  - \lambda \sum_{i=1}^2 V_i \right] \label{Seff}
\end{align}
with 
\begin{align}
	Q_1 = b_1 + 1/b_1 \, , \quad 	Q_2 = b_2 + 1/b_2 \, , \quad 	Q_3 = b_3 - 1/b_3  \label{Qs}
\end{align}
and
\begin{align}
	V_i =  e^{2 \beta_i \cdot \phi} \, , \quad \beta_1 = (b_1 , 0 , 0) \, , \quad \beta_2 = (0 , b_2 , 0) \, . \label{V12}
\end{align}
The relation between the correlation functions in \eqref{corrrel} is up to a factor that is a function of $z_\nu , y_i$.

We then move to correlation functions of vertex operators of the form \eqref{mbasis}.
As in \cite{Hikida:2007tq}, the $\mu_\nu$-integration in \eqref{mbasis} can be mapped to the $y_i$-integration, and the extra insertions of vertex operators at $y_i$ can be interpreted as an interaction term. Thus we can write 
\begin{align}
	\left \langle \prod_{\nu =1}^N \Psi_\nu (z_\nu) \right \rangle 
	= 	\left \langle \prod_{\nu =1}^N \tilde  V_\nu (z_\nu) \right \rangle  \, . \label{corrmap}
\end{align}
The relative factor  is canceled with the Jacobian due to the change of variables from $\mu_\nu$ to $y_i$, see \cite{Ribault:2005wp,Ribault:2005ms,Hikida:2008pe}.
The action for the right hand side is given by \eqref{Seff} but now there are three interaction terms as  
\begin{align}
V_i =  e^{2 \beta_i \cdot \phi} \, ,   \quad \beta_1 = \left(\frac{1}{b_1} , 0 , 0 \right) \, , \quad \beta_2 = \left(0 , \frac{1}{b_2} , 0 \right) \, , \quad \beta_3 = \left(- \frac{1}{2 b_1} , - \frac{1}{2 b_2}  ,  \frac{1}{2 b_3} \right) \, . \label{V123}
\end{align}
For $V_1,V_2$, we have performed the self-duality of Liouville field theory.
As explained in  \cite{Hikida:2008pe}, we may perform reflection relations to $V_1,V_2$ with respect to $V_3$, which lead to
\begin{align}
	\beta_1 = \left(\frac{k_1 -1/2}{b_1},\frac{ k_1 - 3/2}{ b_2},\frac{3/2 - k_1 }{b_3}\right)\, , \quad
	\beta_2 =  \left(\frac{k_2 -3/2}{b_1},\frac{ k_2 - 1/2}{ b_2},\frac{3/2 - k_2}{b_3}\right)\, .
\end{align}
Computing the Gram matrix, we find
\begin{align}
- 2 \beta_i \beta_j = 
\begin{pmatrix}
	1 & k_1 + k_2 - 3 &  - k_1 + 1 \\
	k_1 + k_2 - 3 & 1 &  - k_2 + 1 \\
	- k_1 + 1 & - k_2 + 1 & 1
\end{pmatrix} \, ,
\end{align}
which reproduces (2.11)-(2.14) in \cite{Feigin:2001yq}.%
\footnote{In order to match the convention, we need to replace $k_i$ by $- k_i$ and set $n=-1$.}
This means that the theory with the action \eqref{Seff} corresponds to the free field realization for the coset \eqref{FScoset} given in \cite{Feigin:2001yq}.
In this way, we have  shown that the $N$-point functions of the coset \eqref{FScoset} can be reduced to
 those of the theory with a $\mathfrak{d}(2,1,- \psi)$-structure.

\section{Large $\mathcal{N}=4$ Liouville from a coset}
\label{sec:largeN4}

In this section, we examine the correlation functions of primary operators in the coset \eqref{cosetY} and relate them to those of the large $\mathcal{N}=4$ super Liouville theory, see  \cite{Creutzig:2019kro}. The factor $Y(k_1,k_2)$ in the numerator  is made of two $\mathfrak{d}(2,1;-\psi)$ with $\psi = 1 - k_1,1 -k_2$ at level one. The large $\mathcal{N}=4$  superconformal algebra has two $\mathfrak{sl}(2)$ current algebras with levels $k_1 ' , k_2 '$ satisfying the relations like \eqref{levelrel}. This duality structure is hidden in $\mathfrak{d}(2,1;-\psi)_1$ as explained in section \ref{sec:Dualities}, see \cite{Creutzig:2017uxh} for more details.
In the next subsection, we find out a description of $D(2,1;-\psi)_1$. Since the action in the WZNW model should be quite complicated, we use a description as a coset of $SL(2|1)$ and free bosons \cite{Creutzig:2017uxh}. 
In subsection \ref{sec:red}, we examine the $N$-point functions of primary operators in the coset \eqref{cosetY} and relate them to those of a different theory. In subsection \ref{sec:free}, we identify the theory as the large $\mathcal{N}=4$ super Liouville theory, which corresponds to the free field realization of large $\mathcal{N}=4$ superconformal algebra in \cite{Ito:1992nq}. In subsection \ref{sec:furtherred}, we further gauge $SU(2)_{k_1 ' + k_2 '}$ subsector and reproduce the theory with a $\mathfrak{d}(2,1;-\psi)$-structure obtained in section \ref{sec:HD}.

\subsection{$D(2,1 ; k-1)_1$ from $SL(2|1)_{k}$}
\label{sec:d21}

The factor $Y(k_1,k_2)$ in the numerator of the coset \eqref{cosetY} is made of two $D(2,1|-\psi)$ at level one.
As a simple description of $D(2,1| - \psi)$ at level one, we consider the coset
\begin{align}
	D(2,1 ; k -1)_1 \simeq \frac{SL(2|1)_{k} \otimes U(1)_{\partial Y} \otimes U(1)_{\partial Z}}{U(1)_J}  \label{dcoset}
\end{align}
as in \cite{Creutzig:2017uxh}. Here we set $\psi = 1 - k$.
We introduce free bosons $Y,Z$ with 
OPEs as
\begin{align}
	Y(z) Y(0) \sim - \frac12 \ln |z|^2 \, , \quad Z(z) Z(0) \sim \frac12 \ln |z|^2 \, .
\end{align}
We will later determine the action of $U(1)$-generator $J$ in the denominator of \eqref{dcoset}.
We have already examined the $SL(2|1)$ WZNW model to some extend in section \ref{sec:coset1}.
It will be convenient to adopt the first order formulation of $SL(2|1)$ WZNW model at the level $k$ given in \eqref{sl21action1} as
\begin{align}
	\begin{aligned}
		S^\text{WZNW}_k [g] &= \frac{1}{2 \pi} \int d^2 z \left[  \bar \partial \phi  \partial \phi   -  \bar \partial \varphi  \partial \varphi  + \beta \bar \partial \gamma + \bar \beta \partial \bar \gamma +  p \bar \partial \theta +  \bar p \partial \bar \theta+  q \bar \partial \eta +  \bar q \partial \bar \eta \right] \\
		&\quad + \frac{\lambda }{2  \pi} \int d^2 z \left[  e^{ b ( \phi  + \varphi  ) } (p + \tfrac12 \beta \eta ) (\bar p + \tfrac12 \bar \beta \bar \eta ) + e^{ b ( \phi  - \varphi  ) }( q + \tfrac12 \beta \theta ) (  \bar q + \tfrac12 \bar \beta \bar \theta )   \right] 
	\end{aligned} 
\end{align}
with $b = 1/\sqrt{k-1}$.  We have slightly changed the notation such as
\begin{align}
\phi_1 \to \phi \, , \quad \phi_2 \to \varphi , \quad p_1 \to p \, , \quad \theta_1 \to \theta \, , \quad p_2 \to q \, , \quad \theta_2 \to \eta 
\end{align}
and similarly for $(\bar p_i , \bar \theta_i)$.
Here we use a bosonization formulation as
\begin{align}
	p = e^{ i g^L} \, , \quad \theta = e^{- i g^L} \, , \quad   	q = e^{ i h^L} \, , \quad \theta = e^{-  i h^L} 
\end{align}
with OPEs
\begin{align}
	g^L (z) g^L (0) \sim - \ln  z \, , \quad 	h^L (z) h^L  (0) \sim - \ln  z \, .
\end{align}
We bosonize $\bar p, \bar \theta , \bar q , \bar \eta$ in terms of $g^R , h^R$ in a similar way and define $g = g^L + g^R, h = h^L + h^R$.

The affine Lie superalgebra $\mathfrak{sl}(2|1)_k$ has obvious subalgebra $\mathfrak{sl}(2)_k$ generated by
$E_1^\pm \equiv \pm E^\pm$ and $H_1 \equiv H $.
There is actually hidden $\mathfrak{sl}(2)_{k'}$ as shown in \cite{Bowcock:1999uy,Creutzig:2017uxh}.
The relation of levels is 
given in \eqref{levelrel}.
Generators of the hidden $\mathfrak{sl}(2)_{k'}$  may be expressed as 
\begin{align}
 E^+_2	= e^{2 Y} F^+ F^- \, , \quad E^-_2 = - e^{- 2Y} G^+ G^- \, , \quad H_2 = \left( \frac{k}{1-k}\right) \left( \partial Y + \frac{1}{k} I\right)\, .
\end{align}
With the additional free boson $Z$, the fermionic generators of $\mathfrak{d}(2,1;k-1)$ can be expressed as
\begin{align}
 F^+e^{Y \pm Z} \, , \quad F^-  e^{Y \pm Z} \, , \quad  G^+e^{- Y \pm Z} \, , \quad G^-  e^{-Y \pm Z} \, .
\end{align}
These generators should survive under the gauging of $U(1)_J$.
This determines the action of $J$ as
\begin{align}
	J =  I + \partial Y = - b^{-1} \partial \varphi + \frac{i}{2} (\partial g - \partial h )  + \partial Y \, .
\end{align}
In order to cancel the contribution, we may introduce a boson $W$ with OPE $W (z) W (0) \sim - \frac12 \ln |z| ^2 $ and use the vertex operators of the form
\begin{align}
	V = e^{- \frac{i}{2} (g + h)}\gamma^{m } \bar \gamma^{ \bar m} \bar e^{2 b j \phi + 2 b  l   \varphi + 2 t  Y + 2 u Z + 2 b ( l + t) W} \, . \label{VcosetY}
\end{align}
It will be convenient in working on the Ramond sector indicated by the first factor.
Moreover, it will be enough to deal with $t=0$ sector for our purpose.

\subsection{Reduction of  a coset}
\label{sec:red}

According to  \cite{Creutzig:2019kro}, the factor $Y(k_1 ,k_2)$ is given by a sum of two  $\mathfrak{d}(2,1|-\psi)$ at level one in the coset description \eqref{cosetY} but removing $Z$-fields. 
In the BRST formulation of the coset \eqref{cosetY}, we use the sum of actions as
\begin{align}
 S = S^1_{k_1} + S^2_{k_2} + S^{\mathfrak{sl}(2)}_{- k_1 - k_2 + 4}[\phi_3 , \gamma_3 , \beta_3] + S_{bc} [b^a,c_a] \, .
\end{align}
For the action $S^1_{k_1}$, we use%
\begin{align}
	&	S^1_{k_1} = \frac{1}{2 \pi} \int d^2 z \left[  \bar \partial \phi_1  \partial \phi_1  -  \bar \partial  \varphi_1  \partial  \varphi_1 + \bar \partial  W_1 \partial W_1   + \beta_1 \bar \partial \gamma_1 + \bar \beta_1 \partial \bar \gamma _1 \right] \nonumber \\
		&+ \frac{1}{2 \pi} \int d^2 z \left[    p _1 \bar \partial \theta _1 +  \bar p _1 \partial \bar \theta _1 +  q_1 \bar \partial \eta_1 +  \bar q_1 \partial \bar \eta_1    \right] \\
		&+ \frac{ \lambda }{2 \pi} \int d^2 z  \left[ e^{ b_1 ( \phi_1  + \varphi_1  ) } (p_1 + \tfrac12 \beta_1 \eta_1 ) (\bar p_1 + \tfrac12 \bar \beta_1 \bar \eta_1 ) +  e^{ b_1 ( \phi_1  - \varphi_1  ) } (q_1 + \tfrac12 \beta_1   \theta_1)( \bar q_1 + \tfrac12 \bar \beta_1 \bar \theta_1)   \right]  \, , \nonumber
\end{align}
where $b_1 = 1/\sqrt{k_1-1}$.  Now that we are interested in the correlation functions among primary operators of the form \eqref{VcosetY} with $t=0$,  we have neglected a free boson $Y_1$.
For $S^2_{k_2}$, we replace the subscript $1$ by $2$.
The action $ S^{\mathfrak{sl}(2)}_{- k_1 - k_2 + 4}[\phi_3 , \gamma_3 , \beta_3] $ is for the $\mathfrak{sl}(2)$ WZNW model as in  \eqref{SL2action} and $S_{bc} [b^a,c_a] $ is for the BRST ghosts as in \eqref{ghostaction}.

The coset can be now analyzed in a way similar to \eqref{FScoset}.
We neglect $\beta_1,\gamma_1$ and $\beta_1$ in the action is replaced by \eqref{beta1r}.
The vertex operators are considered of the form
\begin{align}
	V = \gamma_2^{  m_2} \bar  \gamma_2^{  \bar m_2} \gamma_3^{ m_3} \bar \gamma_3^{  \bar m_3} \left[ \prod_{i=1}^2  e^{- \frac{i}{2} (g_i + h_i) + 2 b_i ( j_i \phi_i +  l_i  \varphi _i  + l_i W_i )} \right]   e^{2 b_3 j_3 \phi_3}\, . \label{Vtemp}
\end{align}
Here we set $b_3 = 1/\sqrt{k_3 - 2}$ with $k_3 = - k_1 - k_2 + 4$.
The conditions corresponding to \eqref{FScond} are
\begin{align}
	\begin{aligned}
	& - j_1 -1/2 - j_2 -1/2 -  m_2  - j_3 -  m_3 + 1  = 0 \, , \\
&	\frac{	( - j_1 - j_2  -1 + m ) 	( 2 + j_1  + j_2    - m ) }{k_1 + k_2 - 2} - \frac{j_3 (1 - j_3)}{k_1 + k_2 - 2} = 0 
	\end{aligned}
\end{align}
with some integer $m$. There are also similar conditions with $\bar m_2 , \bar m_3 , \bar m$.
We may set
\begin{align}
	m_2 = -  j_1   -  j_2  -  j_3   \, , \quad m_3 = 0\, , \quad m = m_2 + 1
\end{align}
such that $\beta_3,\gamma_3$ can be neglected.
We thus consider the vertex operators of the form
\begin{align}
&	\Psi_\nu (z_\nu) = \int \frac{d \mu_\nu^2}{|\mu_\mu|^2} V_\nu (z_\nu) \, , \label{psi}\\
	&V _\nu (z_\nu) = e^{\mu_\nu \gamma_2 - \bar \mu_\nu \bar \gamma_2}  \left[ \prod_{i=1}^2 |\mu_\nu|^{2 j_i + 1} e^{ - \frac{i}{2} (g_i + h_i ) + 2  b_i ( j_i ^\nu \phi_i +  l_i^\nu  \varphi + l_i^\nu W_i ) } \right]  |\mu_\nu|^{ 2 j_3^\nu  - 2}e^{2 b_3 (1 - j_3 ^\nu) \phi_3}\, . \label{V}
\end{align}
We have performed a reflection relation to $\phi_3$-direction.
An identity operator can be inserted as in the previous case. Here,  it is given by
\begin{align}
\mathbbm{1} = v^{(s)} (\xi) e^{s (  \frac{i}{2} (g_1 + h_1) + \phi_1 /b_1 -  \phi_3 /b_3) } (\xi)\, , \label{fermionicid}
\end{align}
where the spectral flow operator $v^{(s)} (\xi)$ restricts the domain of integral over $\beta_2$ such as to have a zero of order $s$ and effectively inserts $e^{s (\frac{i}{2} (g_2 + h_2)  + \phi_2/b_2 )}$ at $w = \xi$ as before.

We first examine the correlation functions of the form
\begin{align}
\begin{aligned}
&	\left \langle  v^{(s)} (\xi) e^{s (  \frac{i}{2} (g_1 + h_1) + \phi_1 /b_1 -  \phi_3 /b_3) } (\xi) \prod_{\nu =1}^N V_\nu (z_\nu) \right \rangle \\
&\quad = 
	\int _s \mathcal{D} g e^{ - S} \prod_{\nu =1}^N V_\nu (z_\nu) e^{s (  \sum_{j=1}^2 \frac{i}{2} (g_j + h_j) + \phi_1 /b_1 + \phi_2 /b_2 -  \phi_3 /b_3) } (\xi) \, .
\end{aligned}
\end{align}
The path integral measure is
\begin{align}
\mathcal{D}g  = \left[\prod_{i=1}^3 \mathcal{D} \phi_i  \right] \left[\prod_{i=1}^2 \mathcal{D} \varphi_i \mathcal{D} p_i \mathcal{D} \theta_i  \mathcal{D} q_i \mathcal{D} \eta_i \mathcal{D} W_i \right]  \mathcal{D} \gamma_2  \mathcal{D} \beta_2 
\end{align}
and the effective action is 
\begin{align}
		S &= \frac{1}{2 \pi} \int d^2 z \left[  \sum_{i=1}^3  \bar \partial \phi_i  \partial \phi_i   +  \frac{1}{4} \sqrt{g} \mathcal{R} b_3 \phi_3 + \sum_{i=1}^2 \left( -  \bar \partial  \varphi_i  \partial  \varphi_i + \bar \partial W_i \partial W_i \right) \right]  \\
		&\quad + \frac{1}{2 \pi} \int d^2 z \left[  \sum_{i=1}^2 (  p _i\bar \partial \theta _i +  \bar p _i \partial \bar \theta _i +  q_i \bar \partial \eta_i +  \bar q_i \partial \bar \eta_i ) + \beta_2 \bar \partial \gamma_2 + \bar \beta_2 \partial \bar \gamma _2 + \lambda  \sum_{i=1}^4 V_i \right] \, . \nonumber 
\end{align}
Here the interaction terms are
\begin{align}
	\begin{aligned}
	&V_1 = e^{ b_1 ( \phi _1  + \varphi _1 ) } (p_1 + \tfrac12 \beta_1 \eta_1 ) (\bar p_1 + \tfrac12 \bar \beta_1 \bar \eta_1 )    \, , \quad
	V_2 = e^{ b_1 ( \phi _1 - \varphi_1  ) } (q_1 + \tfrac12 \beta_1 \theta_1) (  \bar q_1 + \tfrac12 \bar \beta_1 \bar \theta_1 )  \, ,    \\
	&V_3 =  e^{ b_2 ( \phi _2 + \varphi _2 ) }  (p_2 + \tfrac12 \beta_2 \eta_2 ) (\bar p_1 +\tfrac12 \bar \beta_2 \bar \eta_2 )    \, , \quad
V_4 = e^{ b_2 ( \phi_2  - \varphi _2 ) } (q_2 + \tfrac12 \beta_2 \theta_2) (  \bar q_2 + \tfrac12 \bar \beta_2 \bar \theta_2 ) \, .
\end{aligned}
\end{align}
As before, integration over zero mode of $\gamma_2$ gives delta functions for $\mu_\nu$ and integration over non-zero modes of $\gamma_2$  leads to a delta functional for $\beta_2$. After integrating $\beta_2$ out, $\beta_2$ is replaced by a function
\begin{align}
-	\beta_2 (w) = \sum_{\nu =1}^N \frac{\mu_\nu}{w - z_\nu} = u \frac{(w - \xi)^s \prod_{i=1}^{N-2-s } (w - y_i)}{ \prod_{\nu=1}^N (w - z_\nu)} = u \mathcal{B} (w ; z_\nu , y_i) \, .
\end{align}
The restrictions coming from delta functions are
\begin{align}
\sum_{\nu=1}^N \frac{\mu_\nu}{(w - \xi)^a} = 0 
\end{align}
 for $a = 0,1,\ldots, s$. 

In order to remove the function appearing in the action, we shift the fields as
\begin{align}
	\begin{aligned}
&	\phi_1 + \frac{1}{2b_1} \ln |u \mathcal{B}|^2 \to \phi_1 \, , \quad 
	\phi_2 + \frac{1}{2b_2} \ln |u \mathcal{B}|^2 \to \phi_2 \,  , \quad 
	\phi_3- \frac{1}{2b_3} \ln |u \mathcal{B}|^2 \to \phi_3 \, , \\
&	g_i + \frac{ i }{2}\ln |u \mathcal{B}|^2 \to g_i \, , \quad 
    h_i + \frac{i }{2} \ln |u \mathcal{B}|^2 \to h_i \, .
    \end{aligned}
\end{align}
The correlation function becomes
\begin{align}
	\left \langle \prod_{\nu=1}^N V_\nu (z_\nu) \right \rangle 
	 = \left \langle \prod_{\nu =1}^N \tilde V_\nu (z_\nu) \prod_{i=1}^{N-2-s} \tilde V_b (y_i) \right \rangle 
\end{align}
up to a factor that is a function of $z_\nu , y_i$.
The vertex operators are
\begin{align}
	\begin{aligned}
	V _\nu (z_\nu) & = \left[  \prod_{i=1}^2e^{2 b_i ( (j_i ^\nu + 1/2 b_i^2)  \phi_i +   l_i^\nu  \varphi_i  +  l_i ^\nu W_i ) }  \right] e^{2 b_3 (1 - j_3 ^\nu - 1/ 2 b_3^2) \phi_3}
 \end{aligned}
\end{align}
and 
\begin{align}
	\tilde V_b (y_i) = e^{- \frac{i}{2} (g_1 + h_1 + g_2 + h_2) - \phi_1 /b_1  - \phi_2 /b_2 + \phi_3 /b_3} \, .
\end{align}
If we consider $\Psi_\nu (z_\nu)$ in \eqref{psi} instead of $V_\nu (z_\nu)$ in \eqref{V}, then the vertex operators $\tilde V_b (y_i)$ can be regarded as an interaction term as above. 
The correlation functions are now written as
\begin{align}
	\left \langle \prod_{\nu=1}^N \Psi_\nu (z_\nu) \right \rangle 
	= \left \langle \prod_{\nu =1}^N \tilde V_\nu (z_\nu)  \right \rangle \, , 
\end{align}
where the factor is canceled by  the Jacobian due to the change of variables $\mu_\nu$ to $y_i$.
The action for the right hand side becomes
\begin{align}
\label{Laction}
\begin{aligned}
		S &= \frac{1}{2 \pi} \int d^2 z \left[  \sum_{i=1}^3  \bar \partial \phi_i  \partial \phi_i   +  \frac{1}{4} \sqrt{g} \mathcal{R} \left( \sum Q_{\phi_i} \phi_i \right)  + \sum_{i=1}^2 \left( -  \bar \partial  \varphi_i  \partial  \varphi_i + \bar \partial  W_i \partial W_i \right) \right]  \\
		&\quad + \frac{1}{2 \pi} \int d^2 z \left[  \sum_{i=1}^2 (  p _i\bar \partial \theta _i +  \bar p _i \partial \bar \theta _i +  q_i \bar \partial \eta_i +  \bar q_i \partial \bar \eta_i )  + \lambda  \sum_{i=1}^5 V_i \right] \, ,
\end{aligned}
\end{align}
where the interaction terms are
\begin{align}
	\begin{aligned}
	&V_1 = e^{ b_1 ( \phi _1  + \varphi _1 ) } (p_1 - \tfrac12  \eta_1 ) (\bar p_1 + \tfrac12 \bar \eta_1 )    \, , \quad
	V_2 = e^{ b_1 ( \phi _1 - \varphi_1  ) } (q_1 - \tfrac12  \theta_1) (  \bar q_1 + \tfrac12  \bar \theta_1 )  \, ,    \\
	&V_3 =  e^{ b_2 ( \phi _2 + \varphi _2 ) }  (p_2 - \tfrac12  \eta_2 ) (\bar p_1 + \tfrac12   \bar \eta_2 )    \, , \quad
V_4 = e^{ b_2 ( \phi_2  - \varphi _2 ) } (q_2 - \tfrac12  \theta_2) (  \bar q_2 + \tfrac12 \bar \theta_2 ) \, , \\
&V_5= e^{- \frac{i}{2} (g_1 + h_1 + g_2 + h_2) - \phi_1 /b_1  - \phi_2 /b_2 + \phi_3 /b_3} \, . 
\end{aligned} \label{Lint}
\end{align}
The background charges are
\begin{align}
	Q_{\phi_1} = \frac{1}{b_1} \, , \quad Q_{\phi_2} = \frac{1}{b_2} \, , \quad Q_{\phi_3} = b_3 - \frac{1}{b_3} 
 \label{bg}
\end{align}
and the conformal weights of $p_i, \theta_i , q_i , \eta_i$ are $1/2$ and similarly for  $\bar p_i, \bar \theta_i , \bar q_i , \bar \eta_i$.

\subsection{Large $\mathcal{N}=4$ super Liouville theory}
\label{sec:free}

In the previous subsection, we have rewritten the correlation functions of the coset \eqref{cosetY} in terms of a different theory with the action \eqref{Laction}. In this subsection, we identify the theory as the $\mathcal{N}=4$ super Liouville theory.
Let us first observe that if we change the fermionic variables as
\begin{align}
\tfrac12 \theta_i +  q_i  \to \theta_i \, , \quad  p_i + \tfrac12 \eta_i  \to p_i \, , \quad 
\tfrac12 \eta_i -  p_i  \to \eta_i \, , \quad  q_i - \tfrac12 \theta_i  \to q_i \, , 
\end{align}
then the theory with interaction terms $V_1,V_2$ (or $V_3,V_4$) can be identified with those of the $\mathcal{N}=2$ super Liouville theory. Therefore, treating an interaction term $V_5$ perturbatively, the theory can be regarded as a sum of two $\mathcal{N}=2$ super Liouville theories with parameters $k_i -1$.

It is known that $\mathcal{N}=2$ super Liouville theory is dual to a super cigar model \cite{Hori:2001ax,Creutzig:2010bt}
\begin{align}
\frac{SL(2)_{k'_i +1} \otimes SO(2)_1}{U(1)} \, ,  \label{scigar2}
\end{align}
where
$
k'_i  - 1 = 1/(k_i - 1)
$ as in  \eqref{levelrel}.
We apply the fermionic FZZ-duality to the two  $\mathcal{N}=2$ super Liouville theories.
Then the correlation functions of the coset \eqref{cosetY} become
\begin{align}
	\left \langle \prod_{\nu=1}^N \Psi_\nu (z_\nu) \right \rangle 
	= \left \langle \prod_{\nu =1}^N \tilde \Psi_\nu (z_\nu)  \right \rangle \, ,
\end{align}
where the right hand side is evaluated with the action
\begin{align}
\begin{aligned}
		S &= \frac{1}{2 \pi} \int d^2 z \left[  \sum_{i=1}^3  \bar \partial \phi_i  \partial \phi_i   +  \frac{1}{4} \sqrt{g} \mathcal{R} \left( \sum Q_{\phi_i} \phi_i \right)  + \sum_{i=1}^2 \left( -  \bar \partial  \varphi_i  \partial  \varphi_i + \bar \partial  W_i \partial W_i \right) \right]  \\
		&\quad + \frac{1}{2 \pi} \int d^2 z \left[  \sum_{i=1}^2 (  p _i\bar \partial \theta _i +  \bar p _i \partial \bar \theta _i +  q_i \bar \partial \eta_i +  \bar q_i \partial \bar \eta_i  + \beta_i \bar \partial \gamma_i + \bar \beta_i \partial \bar \gamma _i)  + \lambda  \sum_{i=1}^3 V_i \right] \,  .
\end{aligned}
\end{align}
The interaction terms are
\begin{align}
	V_1 = e^{ 2 \phi_1 /b_1   } \beta_1 \bar \beta_1  \, , \quad 
	V_2 =  e^{ 2  \phi _2 /b_2   } \beta_2 \bar \beta_2  \label{int}
\end{align}
and $V_3$ obtained by the duality map from $V_5$ as analyzed later.
The background charges are the same as those in \eqref{bg}.
The vertex operators are now
\begin{align}
	\begin{aligned}
	\tilde \Psi_\nu (z_\nu) & = \left[ \prod_{i=1}^2 |\gamma_i| ^{- 2 b_i^2 (j_i^\nu + 1/2b_i^2 +  l_i^\nu)} e^{2 b_i ( (j_i ^\nu + 1/2 b_i^2)  \phi_i +   l_i^\nu  \varphi_i  +  l_i^\nu   W_i ) }  \right] e^{2 b_3 (1 - j_3 ^\nu - 1/ 2 b_3^2) \phi_3} \, .
 \end{aligned} \label{Vmid}
\end{align}
Notice that $\varphi_i$ plays the role of $U(1)$-factor in the denominator of $\eqref{scigar2}$.

Since $\varphi_1,\varphi_2$ (and $W_1,W_2$) do not appear in the interaction terms \eqref{int} anymore (see also $V_5$ in \eqref{Lint}), we can integrate them out.
Due to the form of vertex operators \eqref{Vmid}, the contributions cancel out with each other.
Thus we can use the vertex operators 
\begin{align}
	\begin{aligned}
	\tilde \Psi_\nu (z_\nu) & = \left[ \prod_{i=1}^2 |\gamma_i| ^{- 2 b_i^2 (j_i^\nu + 1/2b_i^2 +  l_i^\nu)} e^{2 b_i (j_i ^\nu + 1/2 b_i^2)  \phi_i}  \right] e^{2 b_3 (1 - j_3 ^\nu - 1/ 2 b_3^2) \phi_3} 
 \end{aligned} 
\end{align}
and the effective action
\begin{align}
\begin{aligned}
		S &= \frac{1}{2 \pi} \int d^2 z \left[  \sum_{i=1}^3  \bar \partial \phi_i  \partial \phi_i   +  \frac{1}{4} \sqrt{g} \mathcal{R} \left( \sum Q_{\phi_i} \phi_i \right)   \right]  \\
		&\quad + \frac{1}{2 \pi} \int d^2 z \left[  \sum_{i=1}^2 (  p _i\bar \partial \theta _i +  \bar p _i \partial \bar \theta _i +  q_i \bar \partial \eta_i +  \bar q_i \partial \bar \eta_i  + \beta_i \bar \partial \gamma_i + \bar \beta_i \partial \bar \gamma _i)  + \lambda  \sum_{i=1}^3 V_i \right]  \, , \label{actionN4}
\end{aligned}
\end{align}
where interaction terms and background charges are the same as before.

It is time to think about $V_3$, which involves a spin field. It would be useful to move to another description of fermions as
\begin{align}
\begin{aligned}
&\psi^{\epsilon_1 \epsilon_2 \epsilon_3 \epsilon_4}=  e^{ \frac{i}{2} ( \epsilon_1 g_1 + \epsilon_2 h_1 + \epsilon_3 g_2 + \epsilon_4 h_2)} 
\end{aligned}
\end{align}
by utilizing the triality relation of $SO(8)$ representations. Here $\epsilon_i = \pm$ and the mutually locality requires that  $\prod_{i=1}^4 \epsilon_i = + 1$. We may decouple the half of fermions furthermore and left
\begin{align}
\psi^{\epsilon_1 \epsilon_2} \equiv \psi^{\epsilon_1 \epsilon_1 \epsilon_2 \epsilon_2} \, , \quad
 \psi^{++} (z)\psi^{--} (0) \sim \frac{1}{z} \, , \quad \psi^{+-} (z)\psi^{-+} (0) \sim \frac{1}{z} \, .
\end{align}
Before decoupling the half of fermions, the interaction term $V_3$ is neutral under the $\mathfrak{sl}(2)_{k_i -1}$ currents,
which are the diagonal sum of $\mathfrak{sl}(2)_{k_i+1}$ and $\mathfrak{sl}(2)_{-2}$ made of fermions.
The $\mathfrak{sl}(2)_{k_i+1}$ currents may be expressed as
\begin{align}
J^+_{b,i}  = \beta_i \, , \quad J^3_{b,i}  = b_i  \partial \phi_i +\beta_i    \gamma_i \, , \quad
J^-_{b,i} = \beta_i \gamma_i \gamma_i + 2 b_i \gamma_i \partial \phi_i - (k_i ' +1) \partial \gamma_i  \, .
\end{align}
After decoupling the half of fermions, the currents made of fermions are modified as $\mathfrak{sl}(2)_{-1}$.
We may express them as 
\begin{align}
\begin{aligned}
J_{f,1}^3 = \tfrac12 (\psi^{++} \psi^{--} + \psi^{+-} \psi^{-+} )\, , \quad 
J_{f,1}^+ = \psi^{+-} \psi^{++} \, , \quad J_{f,1}^- = \psi^{-+} \psi^{--} \, , \\
 J_{f,2}^3 = \tfrac12 (\psi^{++} \psi^{--} - \psi^{+-} \psi^{-+} )\, , \quad 
J_{f,2}^+ = \psi^{-+} \psi^{++} \, , \quad J_{f,3}^- = \psi^{+-} \psi^{--} \, .
\end{aligned} \label{fcurrents}
\end{align}
The interaction term $V_3$ can be fixed up to overall factor by the singlet condition with respect to the symmetry of  $\mathfrak{sl}(2)_{k_1}$ and  $\mathfrak{sl}(2)_{k_2}$, which are the diagonal sum of $\mathfrak{sl}(2)_{k_i+1}$ and $\mathfrak{sl}(2)_{-1}$ made of fermions.
The singlet is then given by
\begin{align}
V_3 = |\gamma_1 \gamma_2 \psi^{++} - \gamma_1 \psi^{+-} - \gamma_2 \psi^{-+} - \psi^{--}|^2 e^{- \phi_1/b_1 - \phi_2/b_2 + \phi_3/b_3} \, ,  \label{V3p}
\end{align}
which is the same as the one in (27) of \cite{Ito:1992nq}.

The energy momentum tensor of the theory is given by
\begin{align}
	T = - \sum_{i=1}^3  \left[ \partial \phi_i \partial \phi_i + Q_{\phi_i} \partial^2 \phi_i \right] 
	 + \tfrac12 \sum_{\epsilon = \pm} (\partial \psi^{+,\epsilon}  \psi^{- ,- \epsilon}  - \psi^{+,\epsilon} \partial \psi^{-,-\epsilon} )- \sum_{i=1}^2 \beta_i \partial \gamma_i  \, ,  \label{T}
\end{align}
whose central charge is 
\begin{align}
	c = 9 + 6 \left(\sum_{i=1}^3 Q_i^2 \right) =  - 3 - \frac{6}{k_1 + k_2  - 2} \, . \label{largeN4center}
\end{align}
We except that the theory has the symmetry of the large $\mathcal{N}=4$ superconformal algebra including $\mathfrak{sl}(2)_{k_1' }$ and $\mathfrak{sl}(2)_{k_2' }$ subalgebras with \eqref{levelrel}.
 Indeed we can rewrite as
\begin{align}
	c = \frac{3 (k_1 ' + k_2 ') - 6 k_1 ' k_2 '}{k_1 ' + k_2 ' -2}  \, ,
\end{align}
which is the expected value of central charge, see, e.g., eq. (B.37) of \cite{Gaberdiel:2013vva} (up to $k_i ' \to - \hat k_i^\pm$).
Fermionic currents are constructed such as to commute with interaction terms as well. From this condition, we find
\begin{align}
&b_3^{-1} G^{++} =  b_2^{-1} \partial \phi_2  \psi^{++} + b_1^{-1} \partial \phi_1 \psi^{++}  - b_3^{-1} \partial \phi_3 \psi^{++} + (1-k_1) \beta_1 \psi^{-+}  +  (k_1 -1)  \gamma_1  \beta_1 \psi^{++}  \nonumber \\
& \quad + (k_2 -1) \gamma_2 \beta_2 \psi^{++}  + (1-k_2) \beta_2 \psi^{+-}  + (k_2 -k_1 ) \psi^{+-} \psi^{++} \psi^{-+}  +  (k_1 +k_2 -1) \partial \psi^{++} \, , \nonumber \\
&b_3^{-1} G^{-+} = b_2^{-1} \partial \phi _2 \psi^{-+}  - b_1^{-1} \partial \phi_1 \psi^{-+} + 2 b_1^{-1} \partial \phi _ 1 \gamma_ 1 \psi^{++} - b_3^{-1} \partial \phi _3 \psi^{-+} + (1-k_1) \gamma_1 \beta_1 \psi^{-+} \nonumber \\
& \quad + (k_2-1) \gamma _2 \beta _2 \psi^{-+} + (k_2-1) \beta _2 \psi^{--} + (k_1-1) \gamma_1 \gamma _ 1 \beta _1 \psi ^{++} + (1-2 k_1 ) \partial \gamma _1 \psi^{++}\nonumber   \\
& \quad + (k_1-k_2) \psi^{++} \psi^{-+} \psi^{--} + (k_1+k_2-1) \partial \psi^{-+} \, , \nonumber \\
&b_3^{-1} G^{+-} = 2 b_2^{-1} \partial \phi_2 \gamma_2 \psi^{++} - b_2^{-1} \partial \phi_2  \psi^{+-} + b_1^{-1} \partial \phi_1 \psi^{+-} - b_3^{-1} \partial \phi_3 \psi^{+-} + (k_1 -1) \beta_1 \psi^{--} \\
&\quad  + (k_1 - 1)  \gamma_1 \beta_1 \psi^{+-} + (1-k_2) \gamma_2  \beta_2 \psi^{+-} + (k_2-1) \gamma_2 \gamma_2 \beta_2 \psi^{++} + (1-2 k_2) \partial \gamma_2  \psi^{++} \nonumber \\
&\quad + (k_1-k_2) \psi^{+-} \psi^{++} \psi^{--} + (k_1+k_2-1) \partial \psi^{+-} \, , \nonumber \\
&b_3^{-1} G^{--} = 2 b_2^{-1} \partial \phi _2 \gamma _2 \psi^{-+} + b_2^{-1} \partial \phi _2 \psi^{--} + b_1^{-1} \partial \phi _1 \psi^{-- } + 2 b_1^{-1} \partial \phi _1 \gamma _1 \psi^{+-} + b_3^{-1} \partial \phi_3 \psi^{--} \nonumber  \\
& \quad + (k_1-1) \gamma _1 \beta _1 \psi^{--} + (k_2-1) \gamma _2 \beta _ 2 \psi^{--}+ (k_2-1) \gamma_2 \gamma _2 \beta _2 \psi^{-+} +  (1-2 k_2 ) \partial \gamma _2 \psi^{-+} \nonumber \\
& \quad + (k_1-1) \gamma_1 \gamma _1 \beta_1 \psi^{+-} + (1-2 k_1) \partial \gamma_ 1 \psi^{+-} + (k_1-k_2) \psi^{+-} \psi^{-+} \psi^{--} - (k_1+k_2-1) \partial \psi^{--} \, .\nonumber 
\end{align}
The OPEs among the generators of large $\mathcal{N}=4$ superconformal algebra are summarized in appendix \ref{sec:OPEs}.

\subsection{Further reduction of the coset}
\label{sec:furtherred}

In \cite{Creutzig:2019kro}, it was also shown that the coset algebra of \eqref{FScoset} can be reproduced from a further coset of \eqref{cosetY} given by
\begin{align}
	 \frac{Y(k_1  , k_2)}{ SU(2)_{k_1 ' } \otimes SU(2)_{k_2 ' } \otimes SU(2)_{k_1 + k_2}} \, . \label{cosetcoset}
\end{align}
We describe it as a $SU(2)_{k_1 ' } \otimes SU(2)_{k_2 ' }$ coset of the large $\mathcal{N}=4$ super Liouville theory obtained above. Here we derive correlator correspondences for the duality.
In the BRST formulation, the action for the coset is given by
\begin{align}
S = S^{\mathcal{N}=4}_{k_1 ' k_2 '}  + \sum_{i=1}^2 S^{\mathfrak{sl}(2)}_{- k_i ' - 4} [\tilde \phi_i , \tilde \beta_i , \tilde \gamma_i]+ S_{bc} [b^a , c_a] \, . 
\end{align}
The first action is for the large $\mathcal{N}=4$ super Liouville theory given by \eqref{actionN4} with \eqref{int} and \eqref{V3p}. We bosonize the fermions as
\begin{align}
\psi^{++} = e^{i Y_1^L} \, , \quad \psi^{--} = e^{-i Y_1^L} \, , \quad 
\psi^{+-} = e^{i Y_2^L} \, , \quad \psi^{-+} = e^{-i Y_2^L} 
\end{align}
with $Y^L _i (z) Y^L _j  (0) \sim - \delta_{i,j} \ln z$
and similarly for $\bar \psi^{\epsilon_1 ,\epsilon_2}$ with $Y_i^R$. We further define $Y_i = Y_i^L + Y_i^R$.
The second and third actions are for $\mathfrak{sl}(2)$ WZNW models with levels $k_1 ' . k_2 '$, respectively, given in \eqref{SL2action}. The fourth action is for the BRST ghosts, which can be found in \eqref{ghostaction}.

We apply the first order formulation of the coset as usual. Namely, in the action of large $\mathcal{N}=4$ super Liouville theory, we remove $ \gamma_1 , \gamma_2 $ and replace $\beta_1 , \beta_2$ in the interaction terms $V_1 , V_2$ in \eqref{int} by $J^+_{f,1} , J^+_{f,2}$ given in \eqref{fcurrents}, see, e.g.,  \eqref{betar}. Now the interaction terms becomes
\begin{align}
V_1 = e^{2  \phi_1 /b_1} e^{ i (Y_1 + Y_2)} \, , \quad V_2 = e^{2  \phi_1 /b_2} e^{i (Y_1 - Y_2)} \, , \quad
V_3 = e^{-\phi_1 /b_1  - \phi_2/b_2 + \phi_3 /b_3} e^{-i Y_1} \, .
\end{align}
In order to express the field space orthogonal to that of denominator of the coset \eqref{cosetcoset}, we may redefine
\begin{align}
2 \phi_1 /b_1 + i (Y_1 + Y_2) \to 2 \phi_1 / b_1 ' \, , \quad 2 \phi_1 /b_2 + i (Y_1 - Y_2) \to 2 \phi_2 / b_2 '
\end{align}
with $b_1' = 1/\sqrt{k_1 - 2}$ and $b_2' = 1/\sqrt{k_2 -2}$. The interaction terms are now 
\begin{align}
V_1 = e^{2  \phi_1 /b_1 '} \, , \quad V_2 = e^{2  \phi_1 /b_2 '}  \, , \quad
V_3 = e^{-\phi_1 /b_1 ' - \phi_2/b_2 ' + \phi_3 /b_3}  \, ,
\end{align}
where  reproduces \eqref{V123}. In this way, we have shown that the coset \eqref{cosetcoset} reduces to the theory with a $\mathfrak{d}(2,1;-\psi)$-structure dual to another coset \eqref{FScoset}.

\section{Boundary FZZ-triality}
\label{sec:boundary}

Up to now, we have examined correlation functions on a worldsheet of sphere topology.
In this section, we examine the FZZ-triality in the presence of boundary, i.e, D-branes.
D-branes in cigar model are classified in  \cite{Ribault:2003ss}, and it was shown  that there are D1-branes and D2-branes (and somehow degenerate D0-branes). Boundary actions for D2-branes in sine-Liouville theory can be derived from those for D1-branes in cigar model as in \cite{Creutzig:2010bt}. However, boundary actions for D2-branes in cigar model have not been obtained yet, and due to this fact, boundary actions for D1-branes in sine-Liouville theory cannot be obtained in the way.
In appendix \ref{sec:sl2bdry}, we construct boundary actions in $SL(2|1)$ WZNW model by following the analysis for branes in $OSP(1|2)$ WZNW model in \cite{Creutzig:2010zp}. From them, we can  obtain boundary actions in sine-Liouville theory both for D2-branes and D1-branes. We repeat the same analysis for supersymmetric setup, and our results reproduce boundary actions for B-branes and A-branes in $\mathcal{N}=2$ super Liouville theory found in \cite{Hosomichi:2004ph}. 

In appendix \ref{sec:sl2bdry}, we obtain boundary actions in $SL(2|1)$ WZNW model at the classical level, and
it is straightforward to include quantum corrections in the first order formulation.
Applying the first order formulation to the coset \eqref{trialitycoset}, we can read off boundary actions in sine-Liouville theory.
However, the classical relations for parameters $\mu_i$ receive quantum corrections and hence the values of parameters $\mu_i$ should be modified.
We do not try to determine the parameters $\mu_i$ in this paper, but in principal it could be done by repeating the analysis of \cite{Hosomichi:2004ph}.

In section \ref{sec:coset1}, we obtained the bulk action of sine-Liouville theory from the coset \eqref{trialitycoset} as in
 \eqref{sineaction}.
For D2-branes, the boundary actions can be read off as
\begin{align}
	\begin{aligned}
	S_\text{boundary} =  \frac{1}{ 2 \pi } \int du \left[\hat b  \mathcal{K}  \phi  +  (\mu_1 \sigma^+ + \mu_2  \sigma^- ) e^{\phi ^L/\hat b + i \sqrt{\kappa } \chi ^L}
	+  (\mu_3 \sigma_+ + \mu_4 \sigma^- ) e^{\phi ^L /\hat b - i \sqrt{\kappa} \chi ^L } \right] 
	\end{aligned}  \label{D2ba}
\end{align}
from those for B-branes in $SL(2|1)$ WZNW model in \eqref{boundary2}.%
\footnote{In a quantum treatment, the boundary interaction term $\beta e^{\phi_1}$ can be neglected since it can be generated by the other boundary interaction terms. Similar arguments hold also for the other cases below.}
Here we have decomposed $\phi = \phi^L (z) + \phi^R (\bar z)$ and $\chi = \chi^L(z) + \chi^R(\bar z)$.
Moreover, $\mathcal{K}$ represents the curvature of boundary and 
 the Neumann boundary conditions are assigned both for $\phi$ and $\chi$.
The boundary fermions $\eta , \bar \eta$ are replaced by $2 \times 2$ matrices
\begin{align}
\sigma^+ =
\begin{pmatrix} 0 & 1 \\
0 & 0 
\end{pmatrix} \, , \quad 
\sigma^- =
\begin{pmatrix} 0 & 0 \\
1 & 0 
\end{pmatrix}
\end{align}
due to the Grassmann even property of exponential type potentials.
The boundary actions \eqref{D2ba} reproduces (2.54) in \cite{Creutzig:2010bt}.
For D1-branes, the boundary actions are
\begin{align}
	\begin{aligned}
	S_\text{boundary} =  \frac{1}{ 2 \pi } \int du \left[ \hat b  \mathcal{K}  \phi  +\mu_1   \sigma^+ e^{\phi ^L/\hat b + i \sqrt{\kappa } \chi ^L}
	+  \mu_2 \sigma^- e^{\phi _L /\hat b - i \sqrt{\kappa} \chi ^L } \right] \, , 
	\end{aligned}
\end{align}
which is obtained from those for A-branes in $SL(2|1)$ WZNW model in \eqref{boundary4}.
Here we assign the Neumann boundary condition for $\phi$  and the Dirichlet boundary condition for $\chi$.
The form was already anticipated as in (E.2) of  \cite{Creutzig:2010bt}.

We can consider the coset \eqref{scoset} with an additional complex fermion.
In this case, we have obtained $\mathcal{N}=2$ super Liouville theory with the bulk action \eqref{sLaction}.
The boundary actions for B-branes are
\begin{align}
	S_\text{boundary} & =  \int d u \left[  \eta (\partial + \bar \partial) \bar \eta  + \frac{\hat b }{2 \pi }  \mathcal{K} \phi \right]  \\
	& \quad + \frac{1}{2 \pi } \int du \left[ (\mu_1 \eta + \mu_2 \bar \eta ) e^{ \hat b {}^{-1} (\phi^L + i \chi^L ) + i Y^L } + (\mu_3 \eta + \mu_4 \bar \eta )  e^{ \hat b {}^{-1} (\phi^L - i \chi^L )  - i Y^L} \right]  \,  ,  \nonumber
\end{align}
which come from those for B-branes in $SL(2|1)$ WZNW model in \eqref{boundary2}.
They reproduce (5.8) in \cite{Hosomichi:2004ph} and (3.28) in \cite{Creutzig:2010bt}.
Here we assign the Neumann boundary conditions for $\phi , \chi$ and $Y$. 
The boundary actions for A-brane are
\begin{align}
	\begin{aligned}
	S_\text{boundary} & =  \int d u \left[ \eta (\partial + \bar \partial) \bar \eta + \frac{\hat b }{2 \pi } \mathcal{K}  \phi \right]
\\
	& \quad + \frac{1}{2 \pi  } \int du \left[ \mu_1 \eta  e^{ \hat b {}^{-1} (\phi^L + i \chi^L ) + i  Y^L } +  \mu_2 \bar \eta   e^{ \hat b {}^{-1} (\phi^L - i \chi^L )  - i Y^L} \right]   \, ,
	\end{aligned}
\end{align}
which come from those for A-branes in $SL(2|1)$ WZNW model in \eqref{boundary4}.
They reproduce (5.28) of \cite{Hosomichi:2004ph}. 
Here we assign the Neumann boundary condition for $\phi$ and the Dirichlet boundary conditions for $\chi$ and $Y$.

\subsection*{Acknowledgements}

The work of TC is supported by NSERC Grant Number RES0048511.
The work of YH is supported by JSPS KAKENHI Grant Number 19H01896, 21H04469, 21H05187.

\appendix

\section{Convention for Lie superalgebras}
\label{sec:conv}

In this appendix, we introduce generators of Lie superalgebras, which
 may be expressed in terms of supermatrix of the form
\begin{align}
	M = 
	\begin{pmatrix}
		A & B \\ C & D
	\end{pmatrix} \, .
\end{align}
Here $A,B,C$ and $D$ are $n \times n$, $n \times m$, $m \times n$ and $m \times m$ matrices, respectively.
We call it as $n|m$-dimensional representation of corresponding Lie superalgebra.
The supertrace is defined by
\begin{align}
	\text{str} \, M = \text{tr}A - \text{tr} D 
\end{align}
and the norm is fixed by $\langle X , Y \rangle = \text{str} \, (X Y)$.
See, e.g., \cite{Frappat:1996pb} for more details of Lie superalgebras.

\subsection{Lie superalgebra $\mathfrak{sl}(2|1)$}

\label{sec:convention}

The Lie superalgebra is generated by four bosonic generators $E^{\pm},H,I$ and four fermionic ones $F^\pm ,G^\pm$. 
In the $2|1$-dimensional representation, they are expressed by
\begin{align}
\begin{aligned}
&H = \begin{pmatrix} \frac12 & 0 & 0 \\
0 & - \frac12 & 0 \\
0 & 0 & 0
\end{pmatrix} \, , \quad
I = \begin{pmatrix} \frac12 & 0 & 0 \\
0 &  \frac12 & 0 \\
0 & 0 & 1
\end{pmatrix} \, , \quad
E^+ = \begin{pmatrix} 0 & 1 & 0 \\
0 & 0 & 0 \\
0 & 0 & 0
\end{pmatrix} \, , \quad
E^- = \begin{pmatrix} 0 & 0 & 0 \\
1 & 0 & 0 \\
0 & 0 & 0 
\end{pmatrix} \, , \\
&F^+ = \begin{pmatrix} 0 & 0 & 0 \\
0 & 0 & 0 \\
0 & 1 & 0
\end{pmatrix} \, , \quad
F^- = \begin{pmatrix} 0 & 0 & 1 \\
0 & 0 & 0 \\
0 & 0 & 0 
\end{pmatrix} \, , \quad
G^+ = \begin{pmatrix} 0 & 0 & 0 \\
0 & 0 & 0 \\
1 & 0 & 0
\end{pmatrix} \, , \quad
G^- = \begin{pmatrix} 0 & 0 & 0 \\
0 & 0 & -1 \\
0 & 0 & 0 
\end{pmatrix} \, .
\end{aligned}
\end{align}
Non-trivial (anti-)commutation relations are given by
\begin{align}
\begin{aligned}
& [ H ,E^\pm] = \pm E^\pm \, , \quad [H , F^\pm] = \pm \frac12 F^\pm \, , \quad [H , G^\pm] = \pm \frac12 G^\pm \, ,  \quad [I ,F^\pm] = \frac12 F^\pm \, , \\
& [ I , G^\pm] = - \frac12 G^\pm \, , \quad [E^+ , E^- ] = 2 H \, , \quad [E^\pm , F^\mp] = - F^\pm \, , \quad [E^\pm ,  G^\mp] = - G^\pm \, , \\
& \{ F^\pm ,G^\mp \} = \mp I + H \, , \quad \{ F^\pm , G^\pm\} = \pm E^\pm \, . 
\end{aligned}
\end{align}
Their norms are
\begin{align}
	\begin{aligned}
&\langle H , H \rangle = \frac12 \, , \quad \langle I , I \rangle = - \frac12 \, , \quad
\langle E^+ , E^- \rangle = \langle E^- , E^+ \rangle = 1 \, , \\
&\langle F^+ , G^- \rangle =\langle F^- , G^+   \rangle    = - 1 \, , \quad
\langle G^- , F^+  \rangle = \langle G^+ , F^- \rangle =  1 
   \end{aligned}
\end{align}
and others are zero.

\subsection{Lie superalgebra $\mathfrak{sl}(n|1)$}

The Lie superalgebra $\mathfrak{sl}(n|1)$ is generated by the Cartan directions $H_i,I$ with $i=1,2,\ldots , n-1$ and bosonic generators $J_{i,j}$ with $i \neq j$ and $i , j = 1 , 2 , \ldots ,n$. Moreover, there are fermionic generators $F_i ,G_i$ with $i=1,2,\ldots n$.
The generators may be expressed by matrices in the $n|1$-dimensional representation.
We introduce elementary matrices $(e_{I,J})_{K,L} = \delta_{I,K} \delta_{J,L}$.
The Cartan generators may be expressed as
\begin{align}
	H_i  = e_{i,i} - e_{i+1,i+1} \, , \quad I = \frac{1}{n-1}\left(\sum_{i=1}^n e_{i,i} + n e_{n+1,n+1}\right) 
\end{align}
with $i=1,2,\ldots , n-1$.
Other bosonic generators are
\begin{align}
	J_{i,j} = e_{i,j} \quad (i \neq j) \, .
\end{align}
The non-trivial commutation relations among them are
\begin{align}
	\begin{aligned}
& [ H_i , J_{k,l} ] = (  \delta_{i,k} -  \delta_{i+1,k} - \delta_{i,l}  + \delta_{i+1,l} )  J_{k,l} \, , \\
&	[J_{ij} , J_{k,l} ] = \delta_{j,k} J_{i,l} - \delta_{i,l} J_{k,j} \, .
	\end{aligned}
\end{align}
The fermionic generators are
\begin{align}
	F_i = e_{n+1,i} \, , \quad G_i = e_{i,n+1} 
\end{align}
and the commutation relations
\begin{align}
\begin{aligned}
&	[H_i , F_j] = - \delta_{i,j} F_j + \delta_{i+1 ,j} F_j \, , \quad
	[H_i , G_j] = \delta_{i,j} G_j - \delta_{i+1 ,j} G_j \, , \\
&	[I , F_j] = F_j \, , \quad [I, G_j] = - G_j \, , \quad [J_{i,j} , G_k] = \delta_{j,k} G_{i} \, , \quad  [J_{i,j} , F_k] = - \delta_{i,k} F_{j}
\end{aligned}
\end{align}
might be useful. 
With this definition, non-trivial norms are given by
\begin{align}
\begin{aligned}
&	\langle H_i , H_j \rangle = G^{(n)}_{ij} \, , \quad \langle I , I \rangle = - \frac{n}{n-1} \, , \\
&	\langle J_{i,j} , J_{k,l} \rangle = \delta_{i,l} \delta_{j,k} \, , \quad 
	\langle G_i , F_j \rangle =  - \langle F_i , G_j \rangle =  \delta_{i,j} \, .
\end{aligned}
\end{align}
Here $G^{(n)}_{ij}$ is the Cartan matrix of $\mathfrak{sl}(n)$.

\section{Large $\mathcal{N}=4$ superconformal algebra}
\label{sec:OPEs}

The large $\mathcal{N}=4$ superconformal algebra is generated by the energy momentum tensor $T$, four spin 
$3/2$ generators $G^{\pm \pm}$ and two set of affine $\mathfrak{sl}(2)$ currents $J^a_i$ with $a=\pm ,3$ and $i=1,2$ \cite{Sevrin:1988ew,Schoutens:1988ig}.
We may denote the two levels of $\mathfrak{sl}(2)$ current algebras by $k'_1$ and $k_2'$, then the central charge is given by
\begin{align}
	c = \frac{3 (k_1 ' + k_2 ') - 6 k_1 ' k_2 '}{k_1 ' + k_2 ' -2}  
\end{align}
as in \eqref{largeN4center}. A free field realization of the algebra is given in subsection \ref{sec:free}.
In this appendix, we summarize the OPEs among the generators.

We can check that the energy momentum tensor \eqref{T} satisfies the usual OPE with the central charge \eqref{largeN4center}. Moreover, we can show that the other generators are primary with respect to the energy momentum tensor. The spin one generators satisfy two affine $\mathfrak{sl}(2)$ current algebras with levels $k'_1$ and $k_2'$ almost by construction. With our convention, the spin $3/2$ generators transform under the action of $\mathfrak{sl}(2)$ currents as
\begin{align}
\begin{aligned}
&J_1^3 (z)G^{\pm \epsilon} (0) \sim \frac{\pm \frac12 G^{\pm \epsilon} (0) }{z} \, , \quad 
J_1^\mp (z) G^{\pm \epsilon } (0) \sim  \frac{\pm G^{\mp \epsilon} (0) }{z} \, , \\
&J_2^3 (z)G^{\epsilon \pm} (0) \sim \frac{\pm \frac12 G^{\epsilon \pm} (0) }{z} \, ,  \quad 
J_2^\mp (z) G^{\epsilon  \pm} (0) \sim  \frac{\pm G^{\epsilon \mp} (0) }{z} 
\end{aligned}
\end{align}
with $\epsilon = \pm$.

The most non-trivial OPEs are among spin $3/2$ fermionic generators, which produce composite operators consisting of two $\mathfrak{sl}(2)$ currents. We find that
\begin{align}
&G^{++} (z) G^{++} (0) \sim \frac{- \frac{2}{k_1' + k_2 ' -2}  J^+_1 J^+_2 (0)}{z} \, , \nonumber\\
&G^{++} (z) G^{-+}(0) \sim \frac{\frac{2  k_1 '}{k_1 ' + k_2 ' -2}  J_2^+ (0)}{z^2} + \frac{\frac{k_2 '}{k_1 ' + k_2 ' -2} \partial J_1^+ (0) - \frac{2}{k_1' + k_2 ' -2} J^3_1 J^+_2 (0) }{z} \, , \nonumber\\
&G^{++} (z) G^{+-}(0) \sim \frac{\frac{ 2 k_2 '}{k_1 ' + k_2 ' -2}  J_1^+ (0)}{z^2} + \frac{\frac{ k_2 '}{k_1 ' + k_2 ' -2} \partial J_1^+ (0) - \frac{2}{k_1' + k_2 ' -2} J^+_1 J^3_2 (0) }{z} \, , \nonumber\\
&G^{++} (z) G^{--}(0) \sim \frac{\frac{ 2 k_2 '}{k_1 ' + k_2 ' -2}  J_1^3 (0) + \frac{ 2 k_1 '}{k_1 ' + k_2 ' -2}  J_1^3 (0) }{z^2} + \frac{\frac{ k_2 '}{k_1 ' + k_2 ' -2} \partial J_1^3 (0)  + \frac{ k_2 '}{k_1 ' + k_2 ' -2} \partial J_2^3 (0) + T (0)}{z} \nonumber\\ & \quad + \frac{\frac{1}{k_1' + k_2 ' -2} (J^3_1 J^3_1 (0) - \frac12 J_1^+ J_1^- (0)- \frac12 J_1^- J_1^+ (0) ) }{z}  \nonumber\\ &\quad + \frac{ \frac{1}{k_1' + k_2 ' -2} (J^3_2 J^3_2 (0) - \frac12 J_2^+ J_2^- (0)- \frac12 J_2^- J_2^+ (0) ) -  \frac{2}{k_1' + k_2 ' -2} J_1^3 J_2^3 (0)}{z} \, , \nonumber\\
&G^{-+} (z) G^{-+} (0) \sim \frac{- \frac{2}{k_1' + k_2 ' -2}  J^-_1 J^+_2 (0)}{z} \, ,\\
&G^{-+} (z) G^{+-}(0) \sim \frac{\frac{ 2 k_2 '}{k_1 ' + k_2 ' -2}  J_1^3 (0) - \frac{ 2 k_1 '}{k_1 ' + k_2 ' -2}  J_1^3 (0) }{z^2} + \frac{\frac{ k_2 '}{k_1 ' + k_2 ' -2} \partial J_1^3 (0)  - \frac{ k_1 '}{k_1 ' + k_2 ' -2} \partial J_2^3 (0) - T (0)}{z} \nonumber\\ & \quad + \frac{-\frac{1}{k_1' + k_2 ' -2} (J^3_1 J^3_1 (0) - \frac12 J_1^+ J_1^- (0)- \frac12 J_1^- J_1^+ (0) ) }{z} \nonumber \\ &\quad +\frac{- \frac{1}{k_1' + k_2 ' -2} (J^3_2 J^3_2 (0) - \frac12 J_2^+ J_2^- (0)- \frac12 J_2^- J_2^+ (0) ) -  \frac{2}{k_1' + k_2 ' -2} J_1^3 J_2^3 (0)}{z} \, , \nonumber \\
&G^{+-} (z) G^{+-} (0) \sim \frac{- \frac{2}{k_1' + k_2 ' -2}  J^+_1 J^-_2 (0)}{z} \, , \nonumber\\
&G^{-+} (z) G^{--} (0) \sim \frac{\frac{ 2 k_2 '}{k_1 ' + k_2 ' -2}  J_1^- (0)}{z^2} + \frac{\frac{ k_2 '}{k_1 ' + k_2 ' -2} \partial J_1^- (0) - \frac{2}{k_1' + k_2 ' -2} J^-_1 J^3_2 (0) }{z} \, ,  \nonumber  \\
&G^{+-} (z) G^{--} (0) \sim \frac{\frac{ 2 k_1 '}{k_1 ' + k_2 ' -2}  J_2^- (0)}{z^2} + \frac{\frac{ k_1 '}{k_1 ' + k_2 ' -2} \partial J_2^- (0) - \frac{2}{k_1' + k_2 ' -2} J^3_1 J^-_2 (0) }{z} \, ,\nonumber \\
&G^{--} (z) G^{--} (0) \sim \frac{- \frac{2}{k_1' + k_2 ' -2}  J^-_1 J^-_2 (0)}{z} \, . \nonumber
\end{align}

\section{Boundary $SL(2|1)$ WZNW model}
\label{sec:sl2bdry}

In this appendix, we find boundary actions for  branes in the $SL(2|1)$ WZNW model.
For this, we follow the strategy taken for the $OSP(1|2)$ WZNW model in \cite{Creutzig:2010zp}.
Assigning two different gluing conditions for $\mathfrak{sl}(2|1)$ currents at the boundary, we can construct two types of branes, which will be called as B-branes and A-branes by following usual convention.

As in the case of $OSP(1|2)$ WZNW model, 
it is convenient to start from the first order formulation only for the bosonic fields. This means that we use the classical action
\begin{align}
	\begin{aligned}
		S &= \frac{k}{2 \pi} \int d^2 z \left[ \bar \partial \phi _1 \partial \phi _1  - \bar \partial \phi _2  \partial \phi _2   - e^{ - \phi _1 - \phi _2 } \bar \partial \theta_1  \partial \bar \theta_1 + e^{ - \phi _ 1  + \phi _ 2 } \bar \partial \theta_2 \partial \bar \theta_2   \right] \\
	&\quad	+ \frac{1}{2 \pi} \int d^2 z \left[ \beta \left(\bar \partial \gamma - \tfrac12 (  \theta_2 \bar \partial \theta_1 + \theta_1 \bar \partial \theta_2  )  \right)+ \bar \beta \left(\partial \bar \gamma - \tfrac12 (  \bar \theta_2 \partial \bar \theta_1 +  \bar \theta_1 \partial \bar \theta_2 )  \right) \right ]  \\
	&\quad	- \frac{1}{2 \pi k} \int d^2 z \beta \bar \beta e^{2 \phi_1}\, .
	\end{aligned}
\end{align}
The equations of motion lead to
\begin{align}
\beta = k e^{ - 2 \phi_1} (\partial \bar \gamma - \tfrac12 (\bar \theta_2 \partial \bar \theta_1 + \bar \theta_1 \partial \bar \theta_2)) \, ,  \quad
\bar 	\beta = k e^{ - 2 \phi_1} (\bar \partial \gamma - \tfrac12 ( \theta_2 \bar \partial  \theta_1 +  \theta_1 \bar \partial  \theta_2)) \, . \label{sl21eom}
\end{align}
We express the $\mathfrak{sl}(2|1)$ currents by
\begin{align}
\bar J (\bar z) &= k g^{-1}\bar \partial g \\
&= \bar J^{F^+} F^+ -  \bar J^{G^+} G^+ +   \bar J^{E^+}  E^+ + 2   \bar J^{H}  H - 2 J^{I} I + \bar J^{F^-} F^-  - \bar J^{G^-} G^- +   \bar J^{E^-}  E^- \, .  \nonumber 
\end{align}
With the help of equations of motion in \eqref{sl21eom}, we find
\begin{align}
\begin{aligned}
\bar J^{F^+} &= k e^{- \phi_1 - \phi_2} \bar \partial \theta_1  - \bar \theta_2 \bar \beta \, , \\
\bar J^{G^+} & = - k e^{- \phi_1 + \phi_2} \bar \partial \theta_2 + \bar \theta_1 \bar \beta  \, , \\
\bar J^{E^+} &= \bar \beta \, , \\
 \bar J^{H}&= k \bar \partial \phi_1 +  \bar \gamma \bar \beta  - \tfrac{k}{2} e^{- \phi_1 - \phi_2} \bar \theta_1 \bar \partial \theta_1 - \tfrac{k}{2} e^{- \phi_1 + \phi_2} \bar \theta_2 \bar \partial \theta_2 
\, , \\
  \bar J^{I} & =  - k \bar \partial \phi_2 - \tfrac{k}{2} e^{- \phi_1 - \phi_2} \bar \theta_1 \bar \partial \theta_1 + \tfrac{k}{2}  e^{- \phi_1 + \phi_2 } \bar \theta_2 \bar \partial \theta_2  + \tfrac{1}{2} \bar \theta_1 \bar \theta_2  \bar \beta \, , \\
 \bar J^{F^-} &= k \bar \partial \bar \theta_2 - k \bar \theta_2 (\bar \partial \phi_1  - \bar \partial \phi_2) - \bar \theta_2 \bar \gamma \bar \beta + k e^{- \phi_1 - \phi_2 } \bar \gamma \bar \partial \theta_1 - \tfrac{k}{2}  e^{- \phi_1 - \phi_2} \bar \theta_1 \bar \theta_2  \bar \partial \theta_1  \, , \\
\bar J^{G^-} &=- k \bar \partial \bar \theta_1 + k \bar \theta_1 (\bar \partial \phi_1  + \bar \partial \phi_2) + \bar \theta_1 \bar \gamma \bar \beta  - k e^{- \phi_1 + \phi_2 } \bar \gamma \bar \partial \theta_2  - \tfrac{k}{2} e^{- \phi_1 + \phi_2} \bar \theta_1 \bar \theta_2  \bar \partial \theta_2 \, , \\
\bar J^{E^-}& = \tfrac{k}{2} (\bar \theta_1 \bar \partial \bar \theta_2 + \bar \theta_2 \bar \partial \bar \theta_1  )+ k \bar \partial  \bar \gamma - 2 k \bar \gamma \bar \partial \phi_1 - \bar \gamma \bar \gamma \bar \beta \\
& \quad   + k e^{- \phi_1 - \phi_2} \bar \gamma \bar \theta_1 \bar \partial \theta_1 + k e^{- \phi_1 + \phi_2} \bar \gamma \bar \theta_2 \bar \partial \theta_2 + k \bar \theta_1 \bar \theta_2 \bar \partial \phi_2 \, .
\end{aligned}
\end{align}
We define $J^a$ in an analogous way but with replacing $\bar \partial \theta_i$ by $- \partial \bar \theta_i$, see \eqref{eom}.

\subsection{Boundary actions for B-branes}

We examine a B-brane corresponding to the gluing conditions
\begin{align}
	J^a = \bar J^a
\end{align}
at the boundary in the current convention.
The conditions can be reproduced from the fields satisfying
\begin{align}
&\beta = \bar \beta \, , \quad  \bar \gamma - \gamma = c e^{ \phi _ 1 } + \tfrac12 (\theta_1 \bar \theta_2 + \theta_2 \bar \theta_1) \, ,\nonumber  \\
&k e^{- \phi_1 - \phi_2} ( \bar \partial \theta_1 + \partial \bar \theta_1 ) = \beta (\bar \theta_2 - \theta_2) \, , \quad
k e^{- \phi_1 + \phi_2} ( \bar \partial \theta_2 + \partial \bar \theta_2) = \beta (\bar \theta_1 - \theta_1) \, , \nonumber \\
&k (\bar \partial - \partial) \phi_1 + c \beta e^{\phi _ 1} - \frac{k}{4} \left[ e^{-\phi_1 - \phi_2 } (\bar \theta_1 - \theta_1) (\bar \partial \theta_1 - \partial \bar \theta_1 ) + e^{- \phi_1 + \phi_2} (\bar \theta_2 - \theta_2) (\bar \partial  \theta_2 - \partial \bar \theta_2)\right] = 0 \, , \nonumber \\
& (\bar \partial - \partial) \phi_2 + \frac{1}{4} \left[ e^{-\phi_1 - \phi_2 } (\bar \theta_1 - \theta_1) (\bar \partial \theta_1 - \partial \bar \theta_1 ) - e^{- \phi_1 + \phi_2} (\bar \theta_2 - \theta_2) (\bar \partial \theta_2 - \partial \bar \theta_2)\right] = 0 \, , \\
&2 (\bar \partial \bar \theta_2 - \partial \theta_2) - (\bar \theta_2 - \theta_2) (\bar \partial + \partial) (\phi_1 - \phi_2) - c  e^{- \phi_2} (\partial \bar \theta_1 - \bar \partial \theta_1) = 0 \, ,\nonumber  \\
&2 (\bar \partial \bar \theta_1 - \partial \theta_1) - (\bar \theta_1 - \theta_1) (\bar \partial + \partial) (\phi_1 + \phi_2) - c  e^{ \phi_2} (\partial \bar \theta_2 - \bar \partial \theta_2) = 0 \nonumber 
\end{align}
at the boundary.

We would like to construct boundary action which leads to the above conditions.
Using the last two equations, we obtain
\begin{align}
	\begin{aligned}
		&\frac{4 -c^2}{4} (\bar \theta_1 - \theta_1) (\bar  \partial \theta_1 -  \partial \bar \theta_1) = 
		- (\bar \theta_1 - \theta_1) (\bar \partial + \partial)  (\bar \theta_1 - \theta_1) \\
	&\quad	+ \frac{c}{2} e^{\phi_2} (\bar \theta_1 - \theta_1) (\bar \partial + \partial)  (\bar \theta_2 - \theta_2) 
		- \frac{c}{4} e^{\phi_2} (\bar \theta_1 - \theta_1) (\bar \theta_2 - \theta_2)  (\bar \partial + \partial)  (\phi_1 - \phi_2) \, , \\
		&\frac{4 -c^2}{4} (\bar \theta_2 - \theta_2) (\bar  \partial \theta_2 - \partial \bar \theta_2) = 
- (\bar \theta_2 - \theta_2) (\bar \partial + \partial)  (\bar \theta_2 - \theta_2) \\
&\quad	+ \frac{c}{2} e^{- \phi_2} (\bar \theta_2 - \theta_2) (\bar \partial + \partial)  (\bar \theta_1 - \theta_1) 
+ \frac{c}{4} e^{-\phi_2} (\bar \theta_1 - \theta_1) (\bar \theta_2 - \theta_2)  (\bar \partial + \partial)  (\phi_1 + \phi_2) \, . 	
\end{aligned}
\end{align}
This implies that the boundary action is given by
\begin{align}
	\begin{aligned}
		&S_\text{boundary} = \frac{1}{2 \pi i} \int d u \beta (\bar \gamma - \gamma - c e^{\phi_1 } - \frac12 ( \theta_1 \bar \theta_2 + \theta_2 \bar \theta_1 ) ) \\
		& \quad +\frac{k}{2 \pi i} \frac{1}{ 4 - c^2 } \int d u  e^{- \phi_1 - \phi_2 }(\bar \theta_1 - \theta_1) (\partial + \bar \partial) (\bar \theta_1 - \theta_1 ) \\
		& \quad +\frac{k}{2 \pi i} \frac{1}{ 4 - c^2 } \int d u  e^{- \phi_1 + \phi_2 }(\bar \theta_2 - \theta_2) (\partial + \bar \partial) (\bar \theta_2 - \theta_2 )\\
		& \quad -\frac{k}{2 \pi i} \frac{1}{ 4 - c^2 } \frac{c}{2} \int d u  e^{- (\phi_1 + \phi_2)/2 }(\bar \theta_1 - \theta_1) (\partial + \bar \partial) \left[e^{- (\phi_1 - \phi_2)/2 }(\bar \theta_2 - \theta_2 ) \right]    \, .
	\end{aligned}
\end{align}

We move to the first order formulation both for the bosonic and fermionic fields by introducing boundary fermions.
The bulk action is given by \eqref{1stcla} and the boundary action is
\begin{align}
	&S_\text{boundary} = \frac{1}{2 \pi i} \int d u 
	\left[ \beta (\bar \gamma - \gamma - c e^{\phi_1}  )
	- (\bar \theta_1 - \theta_1) p_1 - (\bar \theta_2 - \theta_2) p_2 \right] \nonumber \\
	& \quad + \frac{k}{ 2 \pi i } \int d u \eta (\partial + \bar \partial) \bar \eta + \frac{1}{2 \pi i } \int du (\mu_1 \eta + \mu_2 \bar \eta ) e^{(\phi_1 + \phi_2)/2} (\tfrac{1}{4} \beta (\theta_2 + \bar \theta_2) + p_1)   \label{boundary1}\\
	& \quad + \frac{1}{2 \pi i } \int du  (\mu_3 \eta + \mu_4 \bar \eta ) e^{(\phi_1 - \phi_2)/2} (\tfrac{1}{4} \beta (\theta_1 + \bar \theta_1) + p_2)  \, . \nonumber 
\end{align}
Here the parameters $\mu_1,\mu_2 ,\mu_3 , \mu_4$ should satisfy
\begin{align}
	\mu_3 \mu_2 - \mu_1 \mu_4 = \frac{2 (4 - c^2)}{k c} \, , \quad \mu_3 \mu_4 = \mu_1 \mu_2 = \frac{4 (c^2 -4 )}{k c^2} \, . 
\end{align}
At the boundary we assign
\begin{align}
	\beta = \bar \beta \, , \quad p_1 = \bar p_1 \, , \quad p_2 = \bar p_2 \, .
\end{align}
The first line of \eqref{boundary1} may be included in the bulk action as
\begin{align}
	\begin{aligned}
		S &= \frac{1}{2 \pi} \int d^2 z \left[ k \bar \partial \phi _1 \partial \phi _1  - k \bar \partial \phi _2 \partial \phi _2  - \gamma \bar \partial \beta - \bar \gamma \partial \bar \beta + \sum_{a=1}^2 (  \theta_a \bar \partial  p_a +  \bar \theta_a  \partial \bar  p_a )   -  \frac{1}{k} e^{ 2 \phi _ 1 } \beta \bar \beta  
		\right] \\
		&\quad + \frac{1}{2 k \pi} \int d^2 z \left[  e^{\phi _1 + \phi _ 2 } (p_1 + \tfrac12 \beta \theta_2 ) (\bar p_1 + \tfrac12 \bar \beta \bar \theta_2 ) - e^{\phi _ 1 - \phi _ 2} (p_2 + \tfrac12 \beta \theta_1 ) (\bar p_2 + \tfrac12 \bar \beta \bar \theta_1 ) \right] \, ,
	\end{aligned} \label{1stcla2}
\end{align}
see, e.g., \cite{Fateev:2007wk,Creutzig:2008ek} as well.
Then the boundary action becomes
\begin{align}
	\begin{aligned}
	S_\text{boundary} & =\frac{k}{ 2 \pi i } \int d u \eta (\partial + \bar \partial) \bar \eta 
	- \frac{c}{ 2 \pi i } \int d u  \beta e^{\phi_1} \\
	& \quad + \frac{1}{2 \pi i } \int du (\mu_1 \eta + \mu_2 \bar \eta ) e^{(\phi_1 + \phi_2)/2} (\tfrac{1}{4} \beta (\theta_2 + \bar \theta_2) + p_1) \\
	& \quad + \frac{1}{2 \pi i } \int du  (\mu_3 \eta + \mu_4 \bar \eta ) e^{(\phi_1 - \phi_2)/2} (\tfrac{1}{4} \beta (\theta_1 + \bar \theta_1) + p_2)  \, .
	\end{aligned} \label{boundary2}
\end{align}

\subsection{Boundary actions for A-branes}

In this subsection, we examine an A-brane corresponding to the gluing conditions
\begin{align}
	\begin{aligned}
	&J^H = \bar J^H \, , \quad J^I = - \bar J^I \, , \quad J^{F^+} = - \bar J^{G^+} \, , \quad J^{G^+} = - \bar J^{F^+} \, , \\
	 &J^{F^-} = - \bar J^{G^-} \, , \quad J^{G^-} = - \bar J^{F^-} \, , \quad J^{E^+} = \bar J^{E^+} \, , \quad J^{E^-} = \bar J^{E^-} 
	 \end{aligned}
\end{align}
at the boundary. The conditions can be reproduced from the fields satisfying 
\begin{align}
 \begin{aligned}
&	\beta = \bar \beta \, , \quad \bar \gamma - \gamma = c e^{\phi_1} + \tfrac12 (\theta_1 \bar \theta_1 + \theta_2 \bar \theta_2) \, , \\
& k e^{- \phi_1}(e^{\phi_2} \partial \bar \theta_2 + e^{- \phi_2} \bar \partial \theta_1) = \beta (\bar \theta_2 - \theta_1) \, , \quad 	
k e^{- \phi_1}(e^{-\phi_2} \partial \bar \theta_1 + e^{ \phi_2} \bar \partial \theta_2) = \beta (\bar \theta_1 - \theta_2) \, ,  \\
& k (\bar \partial - \partial) \phi_1 + c \beta e^{\phi_1} \\
& \quad   - \frac{k }{4} e^{- \phi_1}
\left[ (\theta_2 - \bar \theta_1) (e^{\phi_2} \partial \bar \theta_2 - e^{-\phi_2} \bar \partial \theta_1) +  (\theta_1 - \bar \theta_2) (e^{-\phi_2} \partial \bar \theta_1 - e^{\phi_2} \bar \partial \theta_2) \right] = 0 \, , \\
& k (\bar \partial + \partial) \phi_2  + \frac{k }{4} e^{- \phi_1}
\left[ (\theta_2 - \bar \theta_1) (e^{\phi_2} \partial \bar \theta_2 - e^{-\phi_2} \bar \partial \theta_1) -  (\theta_1 - \bar \theta_2) (e^{-\phi_2} \partial \bar \theta_1 - e^{\phi_2} \bar \partial \theta_2) \right] = 0 \, ,  \\
&2 (\bar \partial \bar \theta_1 - \partial \theta_2) - (\bar \theta_1 - \theta_2) \left[ ( \bar \partial + \partial) \phi_1 + (\bar \partial - \partial) \phi_2 \right] - c  ( e^{\phi_2} \bar \partial \theta_2 -  e^{-\phi_2} \partial \bar \theta_1)  = 0 \, , \\
&2 (\bar \partial \bar \theta_2 - \partial \theta_1) - (\bar \theta_2 - \theta_1) \left[ ( \bar \partial + \partial) \phi_1 - (\bar \partial - \partial) \phi_2 \right] - c  ( e^{-\phi_2} \bar \partial \theta_1 -  e^{\phi_2} \partial \bar \theta_2)  = 0 
\end{aligned}	
\end{align}
at the boundary.

We would like to realize these equations from a theory consisting with free kinetic terms and interaction terms. For free parts, we should assign Dirichlet boundary condition for $\phi_2$. In order to realize this, we may set
\begin{align}
	\phi_2 = \phi^L_2 (z) + \phi^R_2 (\bar z) = 0
\end{align}
at the boundary. We further introduce a dual variable $	\tilde \phi_2 = \phi^L_2 (z) - \phi^R_2 (\bar z) $.
The above equations are now given by
\begin{align}
	\begin{aligned}
		&	\beta = \bar \beta \, , \quad \bar \gamma - \gamma = c e^{\phi_1} + \tfrac12 (\theta_1 \bar \theta_1 + \theta_2 \bar \theta_2) \, , \\
		& k e^{- \phi_1}(\partial \bar \theta_2 +  \bar \partial \theta_1) = \beta (\bar \theta_2 - \theta_1) \, , \quad 	
		k e^{- \phi_1}(\partial \bar \theta_1 +  \bar \partial \theta_2) = \beta (\bar \theta_1 - \theta_2) \, ,  \\
		& k (\bar \partial - \partial) \phi_1 + c \beta e^{\phi_1}   - \frac{k }{4} e^{- \phi_1}
		\left[ (\theta_2 - \bar \theta_1) ( \partial \bar \theta_2 -  \bar \partial \theta_1) +  (\theta_1 - \bar \theta_2) ( \partial \bar \theta_1 - \bar \partial \theta_2) \right] = 0 \, , \\
		& k (\bar \partial - \partial) \tilde \phi_2  - \frac{k }{4} e^{- \phi_1}
		\left[ (\theta_2 - \bar \theta_1) ( \partial \bar \theta_2 -  \bar \partial \theta_1) -  (\theta_1 - \bar \theta_2) ( \partial \bar \theta_1 - \bar \partial \theta_2) \right] = 0 \, ,  \\
		&2 (\bar \partial \bar \theta_1 - \partial \theta_2) - (\bar \theta_1 - \theta_2) ( \bar \partial + \partial) ( \phi_1 - \tilde \phi_2 )  - c  (\bar \partial \theta_2 -  \partial \bar \theta_1)  = 0 \, , \\
		&2 (\bar \partial \bar \theta_2 - \partial \theta_1) - (\bar \theta_2 - \theta_1)  ( \bar \partial + \partial) (\phi_1 + \tilde \phi_2 )  - c  ( \bar \partial \theta_1 -  \partial \bar \theta_2)  = 0 \, .
	\end{aligned}	
\end{align}
From the last two equations we find
\begin{align}
	\begin{aligned}
	&	2 (\theta_2 - \bar \theta_1) (\bar \partial + \partial) (\bar \theta_2 - \theta_1 ) 
		- (\theta_2 - \bar \theta_1) (\bar \theta_2 - \theta_1) (\bar \partial + \partial) (\phi_1 + \tilde \phi_2) 
		\\& \qquad \qquad \qquad \qquad \qquad \qquad  - (c - 2) (\theta_2 - \bar \theta_1) (\bar \partial \theta_1 - \partial \bar \theta_2) = 0 \, , \\
	&	2 (\theta_1 - \bar \theta_2) (\bar \partial + \partial) (\bar \theta_1 - \theta_2 ) 
- (\theta_1 - \bar \theta_2) (\bar \theta_1 - \theta_2) (\bar \partial + \partial) (\phi_1 - \tilde \phi_2) 
 	\\& \qquad \qquad \qquad \qquad \qquad \qquad  - (c - 2) (\theta_1 - \bar \theta_2) (\bar \partial \theta_2 - \partial \bar \theta_1) = 0 \, . 
\end{aligned}
\end{align}
This implies that the boundary action is
\begin{align}
	\begin{aligned}
		&S_\text{boundary} = \frac{1}{2 \pi i} \int d u \beta (\bar \gamma - \gamma - c e^{\phi} - \frac12 ( \theta_1 \bar \theta_1 + \theta_2 \bar \theta_2 ) ) \\
		& \quad + \frac{k}{2 \pi i} \frac{1}{c-2}  \int d u  e^{- (\phi_1 + \tilde \phi_2)/2 }(\bar \theta_1 - \theta_2) (\partial + \bar \partial) \left[e^{- (\phi_1 - \tilde \phi_2)/2 }(\theta_1 - \bar \theta_2) \right]    \, .
	\end{aligned}
\end{align}

We move to the first order formulation also for the fermionic fields. With the bulk action \eqref{1stcla},  the boundary action is 
\begin{align}
 \label{boundary3}
	&S_\text{boundary} = \frac{1}{2 \pi i} \int d u 
	\left[ \beta (\bar \gamma - \gamma - c e^{\phi_1}  )
	- (\bar \theta_2 - \theta_1) p_1 - (\bar \theta_1 - \theta_2) p_2 \right] \nonumber \\
	& \quad + \frac{k}{ 2 \pi i } \int d u \eta (\partial + \bar \partial) \bar  \eta + \frac{\mu_1}{2 \pi i } \int du  \eta e^{(\phi_1 + \tilde \phi_2)/2} (\tfrac{1}{4} \beta (\theta_2 + \bar \theta_1) + p_1)   \\
	& \quad + \frac{  \mu_2 }{2 \pi i } \int du \bar \eta  e^{(\phi_1 - \tilde \phi_2)/2} (\tfrac{1}{4} \beta (\theta_1 + \bar \theta_2) + p_2)  \, . \nonumber 
\end{align}
Here the parameters should satisfy
\begin{align}
	\mu_1 \mu_2 = \frac{1}{c-2} \, .
\end{align}
The boundary conditions are now assigned as
\begin{align}
	\beta = \bar \beta \, , \quad \phi_2 = 0 \, , \quad p_1 = \bar p_2 \, , \quad p_2 = \bar p_1 \, .
\end{align}
In terms of the bulk action \eqref{1stcla2}, the boundary action is 
\begin{align}
	&S_\text{boundary} =  \frac{k}{ 2 \pi i } \int d u  \eta (\partial + \bar \partial)  \bar \eta  - \frac{c}{2 \pi i} \int d u \beta  e^{\phi_1 } \label{boundary4} \\
	& \quad + \frac{\mu_1}{2 \pi i } \int du  \eta e^{(\phi_1 + \tilde \phi_2)/2} (\tfrac{1}{4} \beta (\theta_2 + \bar \theta_1) + p_1)   + \frac{  \mu_2 }{2 \pi i } \int du \bar \eta  e^{(\phi_1 - \tilde \phi_2)/2} (\tfrac{1}{4} \beta (\theta_1 + \bar \theta_2) + p_2)  \, . \nonumber 
\end{align}

\section{Generalized FZZ-triality}
\label{sec:gtriality}

In section \ref{sec:coset1}, the coset \eqref{trialitycoset} has been examined and its correlation functions of primary operators have been shown to match with those of sine-Liouville theory or cigar model described by \eqref{cigar}.
In succeeding sections, it was shown how the properties are useful to derive other dualities, such as the one between the coset \eqref{cosetY} and large $\mathcal{N}=4$ super Liouville theory.
Recall that many trialities were conjectured by Gaiotto-Rap\v{c}\'ak \cite{Gaiotto:2017euk} and in particular a series of generalized FZZ-dualities were derived in \cite{Creutzig:2021cyl}. In this appendix we examine its relation to a coset of the type \eqref{hrcoset},
whose symmetry algebra is  $Y_{n,1,0}$-algebra in the notation of \cite{Gaiotto:2017euk}.
We name the relation as generalized FZZ-triality and expect it to be useful for other dualities as in the case of the original FZZ-triality.

\subsection{Bosonic triality}

We start from a free field realization of affine Lie algebra $\mathfrak{sl}(n)$, see, e.g., \cite{Kuwahara:1989xy}.
We introduce $n$ free bosons $\phi_a$ and $n(n-1)/2$ pairs of $(\beta_{i,j},\gamma_{i,j})$-systems with $i  > j$, where the weights of $(\beta_{i,j},\gamma_{i,j})$ are $(1,0)$.
A linear combination of $n$ free bosons $\phi_a$ decouples from the rest.
The non-trivial OPEs are
\begin{align}
	\phi_a (z) \phi_b (0) \sim - \delta_{a,b} \ln z \, , \quad \gamma_{i,j} (z) \beta_{k,l} (0) \sim \frac{\delta_{i,k} \delta_{j,l}}{z} \, .
\end{align}
The currents $J^{\mathfrak{sl}(n)}_{i,j}$ with $i < j$ are given by
\begin{align}
	J_{i,j} ^{\mathfrak{sl}(n)}= \beta_{j,i} - \sum_{l=j+1}^n \gamma_{l , j } \beta_{l , i}
 \end{align}
and the Cartan subalgebra is generated by
\begin{align}
	H^{\mathfrak{sl}(n)}_a =  \hat H^{\mathfrak{sl}(n)}_a - \hat H^{\mathfrak{sl}(n)}_{a+1} 
\end{align}
with
\begin{align}	
\hat H^{\mathfrak{sl}(n)}_a =  \sqrt{k-n} \partial \phi_a - \sum_{l=1}^{a-1} \gamma_{a , l} \beta_{a , l} + \sum_{l=a+1
}^n \gamma_{l , a} \beta_{l , a} \, .
\end{align}
The other generators $J^{\mathfrak{sl}(n)}_{i,j}$ $(i>j)$ can be fixed through the OPEs with these currents.

We then construct a free field realization of affine Lie superalgebra $\mathfrak{sl}(n|1)$.
We prepare an additional free boson $\varphi$ and $n$ pairs of free fermions $(p_j,\theta_j)$ with $j=1,2,\ldots,n$ such that
\begin{align}
	\varphi (z) \varphi (0) \sim \ln z \, , \quad 	p_i (z) \theta_j (0) \sim \frac{\delta_{i,j}}{z} \, .
\end{align}
The weights of $(p_j,\theta_j)$  are $(1,0)$, respectively.
We look for a free field realization such that $J_{i,j} = J^{\mathfrak{sl}(n)}_{i,j}$ for $i < j$.
From the consistency with $J_{i,j} = J^{\mathfrak{sl}(n)}_ {i,j}$, we find
\begin{align}
	F_i = p_i + \sum_{l = i}^{n} \gamma_{l,i} p_l \, .
\end{align}
The Cartan generators are similarly obtained as
\begin{align}
	H_a =  \hat H_a - \hat H_{a+1} \, , \quad 
\hat H_a =  \sqrt{k-n+1} \partial \phi_a - \sum_{l=1}^{a-1} \gamma_{a , l} \beta_{a , l} + \sum_{l=a+1
}^n \gamma_{l , a} \beta_{l , a} - p_a \theta_a
\end{align}
and
\begin{align}
	I= \frac{\sqrt{k - n + 1 }}{n-1} \left( n \partial \varphi +  \sum_{a=1}^n \partial \phi_a \right) + \sum_{j=1}^n p_j \theta_j \, .
\end{align}

From the conditions to commute with these currents, we find out screening operators as
\begin{align}
	Q_i = \int dz V_i(z)
\end{align}
with
\begin{align}
	V_l = \left( \beta_{l+1 , l} - \sum_{j=1}^{l-1} \beta_{l+1 , j} \gamma_{l , j} - p_{l+1} \theta_{l} \right) e^{\frac{1}{\sqrt{k-n+1}} (\phi_l - \phi_{l+1})}  
\end{align}
for $l=1,2,\ldots,n-1$ and
\begin{align}
	V_0 = p_1 e^{- \frac{1}{\sqrt{k-n+1}} ( \varphi + \phi_1 ) } \, .
\end{align}

We then move to find a free field realization of the symmetry algebra of the coset \eqref{hrcoset} by applying the method reviewed in section \ref{sec:1storder}.
It is convenient to bosonize the free fermions as
\begin{align}
	p_i = e^{i X_i} \, , \quad \theta_i = e^{ - i X_i} \, , \quad X_i (z) X_j (0) \sim - \ln z \, .
\end{align}
We then introduce new bosons by
\begin{align}
\hat H^a = \sqrt{k-n} \partial \hat \phi_a - \sum_{l=1}^{a-1} \gamma_{a , l} \beta_{a , l} + \sum_{l=a+1
}^n \gamma_{l , a} \beta_{k , a} \, , \quad 
	I= \sqrt{\frac{nk}{n-1}} \partial  \hat \varphi \, .
\end{align}
The field space of the coset \eqref{hrcoset} should be orthogonal to $\hat \phi_a - \hat \phi_{a+1}$ and $\hat \eta$. Furthermore, $(\beta_{i,j},\gamma_{i,j})$ with all $i > j$ are ignored.
We define new bosons by
\begin{align}
\sqrt{k-n+1} X_a + i  \phi_a  =  - \sqrt{k-n} \hat X_a \, , \quad  \varphi + i \frac{\sqrt{k-n}}{n} \sum_{a=1}^n \hat X_a = - i \sqrt{\frac{k}{n}} \hat \chi \, .
\end{align}
The combinations $\hat x_a = \hat X_a - \hat X_{a +1}$ with $a=1,2,\ldots,n-1$ and $\hat \chi$ are regular with respect to  $\hat \phi_a - \hat \phi_{a+1}$ and $\hat \varphi$.
The screening operators are now written as
\begin{align}
	V_l = - e^{ i \sqrt{\frac{ k - n  }{k - n + 1}} (\hat X_{l} - \hat X_{l+1} )} \, , \quad V_0 = e^{-  i \sqrt{\frac{k - n }{k - n + 1}}  ( \hat X_1  - \frac{1}{n} \sum_{a=1}^n \hat X_a ) + i \sqrt{\frac{k}{n (k - n + 1)}} \hat \chi} \, .
\end{align}
Note that 
\begin{align}
 \hat X_1  - \frac{1}{n} \sum_{a=1}^n \hat X_a = \sum_{l=1}^n G^{(n)1l} \hat x_l \, .
 \end{align}
Moreover, $G^{(n)}_{ij}$ is the Cartan matrix of $\mathfrak{sl}(n)$ and $G^{(n)ij}$ is the inverse matrix of $G^{(n)}_{ij}$.
These are the same as those for $Y_{1,0,n}$-algebra obtained in appendix B of \cite{Creutzig:2021cyl} once the level $k$ is identified with that of \cite{Creutzig:2021cyl} (denoted $\kappa$ here) as
\begin{align}
	\frac{ n - k }{k - n + 1} = \kappa - n \, .
\end{align}

\subsection{Fermionic triality}

In this subsection, we examine the coset
\begin{align}
	\frac{SL(n|1)_k \otimes SO(2)_1}{SL(n)_k \otimes U(1)} \, , 
	\label{sln1scoset}
\end{align}
where $SO(2)_1$ represents a complex fermion $\psi^\pm$ with weight $1/2$.
We may bosonize the fermions as
\begin{align}
	\psi^\pm = e^{ \pm i Y} \,  , \quad Y (z) Y (0) \sim - \ln z \, .
\end{align}
The generators of $\mathfrak{sl}(n)_k$ in the denominator is the same as before.
However, the generator of $\mathfrak{u}(1)$ is shifted as
\begin{align}
	\tilde I = I + \psi^+ \psi^- = I + i \partial Y \, .
\end{align}
In order to generate the orthogonal space, we need another field  $\hat Y$ in addition to $\hat X_a , \hat \chi$.
We may choose the field  as
\begin{align}
 i 	\sqrt{\frac{k n}{n -1} \left(1 + \frac{k n}{n -1}\right)} \hat Y =  \frac{\sqrt{k-n+1}}{n-1} \left( n \varphi + \sum_{a=1}^n \phi_a \right) +i \sum_{i=1}^n X_a - i \frac{k n}{ n - 1} Y \, . 
\end{align}
Further rotating the fields as
\begin{align}
& \sqrt{\frac{k}{n (k - n + 1)}} \hat \chi \to  \sqrt{\frac{k}{n (k - n +1)} - 1 } \hat \chi -  \hat  Y \, , \\
& 	\sqrt{\frac{k}{n (k - n + 1)}} \hat Y \to     \hat \chi +  \sqrt{\frac{k}{n (k - n +1)} - 1} \hat   Y \, ,
\end{align}
the screening operators become
\begin{align}
	V_l = - e^{i \sqrt{\frac{k - n }{k - n + 1}} (\hat X_{l+1} - \hat X_{l} )} \, , \quad V_0 = e^{-  i \sqrt{\frac{k - n }{k - n + 1}}  ( \hat X_1  - \frac{1}{n} \sum_{a=1}^n \hat X_a ) + i \sqrt{\frac{k}{n (k - n +1)} - 1 } \hat \chi -  i \hat  Y} \, .
\end{align}
They reproduce (4.20) with (4.19) in \cite{Creutzig:2021cyl}.


\providecommand{\href}[2]{#2}\begingroup\raggedright\endgroup

\end{document}